\documentclass[fleqn,usenatbib]{mnras}

\usepackage{newtxtext,newtxmath}
\usepackage[T1]{fontenc}

\DeclareRobustCommand{\VAN}[3]{#2}
\let\VANthebibliography\thebibliography
\def\thebibliography{\DeclareRobustCommand{\VAN}[3]{##3}\VANthebibliography}

\usepackage{graphicx}	
\usepackage{xcolor}

\usepackage{amsmath}
\graphicspath{{./}{figures/}}

\title[Galaxy $z$ uncertainties \& impact on dark standard siren $H_0$]{Impact of modelling galaxy redshift uncertainties on the gravitational-wave dark standard siren measurement of the Hubble constant}

\author[C.~Turski et al.]{
Cezary Turski,\textsuperscript{1}\thanks{E-mail: cezary.turski@ugent.be (CT)}
Maciej Bilicki,\textsuperscript{2}
Gergely Dálya,\textsuperscript{1}
Rachel Gray,\textsuperscript{3,4}
and Archisman Ghosh\textsuperscript{1} \\
\textsuperscript{1}{Department of Physics \& Astronomy, Ghent University, Proeftuinstraat 86, 9000 Ghent, Belgium} \\
\textsuperscript{2}{Center for Theoretical Physics, Polish Academy of Sciences, al.~Lotników 32/46, 02-668 Warsaw, Poland} \\
\textsuperscript{3}{Department of Physics \& Astronomy, Queen Mary University of London, Mile End Road, London, E1 4NS, United Kingdom} \\
\textsuperscript{4}{SUPA, University of Glasgow, Glasgow, G12 8QQ, United Kingdom} \\
}

\date{Accepted XXX. Received YYY; in original form ZZZ}

\pubyear{2023}

\begin{document}
\label{firstpage}
\pagerange{\pageref{firstpage}--\pageref{lastpage}}
\maketitle

\begin{abstract}
Gravitational wave science is a new and rapidly expanding field of observational astronomy. Multimessenger observations of the binary neutron star merger GW170817 have provided some iconic results including the first gravitational-wave standard-siren measurement of the Hubble constant, opening up a new way to probe cosmology. The majority of the compact binary sources observed in gravitational waves are however without bright electromagnetic counterparts. In these cases, one can fall back on the ``dark standard siren'' approach to include information statistically from potential host galaxies. For such a measurement, we need to be cautious about all possible sources of systematic errors. In this paper, we begin to study the possible errors coming from the galaxy catalogue sector, and in particular, look into the effect of galaxy redshift uncertainties for the cases where these are photometry-based. We recalculate the dark standard siren Hubble constant using the latest GWTC-3 events and associated galaxy catalogues, with different galaxy redshift uncertainty models, namely, the standard Gaussian, a modified Lorentzian, and no uncertainty at all. We find that not using redshift uncertainties at all can lead to a potential bias comparable with other potential systematic effects previously considered for the GWTC-3 $H_0$ measurement (however still small compared to the overall statistical error in this measurement). The difference between different uncertainty models leads to small differences in the results for the current data; their impact is much smaller than the current statistical errors and other potential sources of systematic errors which have been considered in previous robustness studies.

\end{abstract}


\section{Introduction} \label{sec:intro}

The discovery of gravitational waves (GWs) has opened a new window to astronomical observations \citep{GW150914}. The latest Third Gravitational-Wave Transient Catalogue (GWTC-3) of mergers of black holes (BHs) and neutron stars (NSs) contains 90 events observed by the LIGO-Virgo-KAGRA detector network \citep{LVK_GWcatalog}. Among the many scientific results obtained using these observations is a novel GW measurement of the Hubble constant $H_0$ \citep{LVK_H0}. Future observing runs with increasing detector sensitivities are expected to detect many more such compact binaries. The large number of events expected in the coming years offer the potential to precisely probe the cosmological parameters governing the expansion of the universe. 

The possibility of using GWs to determine the Hubble constant $H_0$ was proposed by \cite{Schutz_H0}, where he pointed out that the luminosity distance of compact binary coalescences can be measured directly from the observed GW data. These sources are thus self-calibrated distance indicators, or `standard sirens'. If a successful electromagnetic (EM) follow-up campaign can identify the host galaxy of the event, $H_0$ can be in principle determined using the distance and the host galaxy's redshift \citep{Schutz_H0,Holz:2005df,MacLeod:2007jd,Nissanke:2009kt,GW170817_H0}. Furthermore, even if an EM counterpart cannot be identified, a statistical analysis using galaxy data from the localization region can be a viable alternative, given a large enough number of observations \citep{Schutz_H0,Del_Pozzo_GW_cosmology,Chen:2017rfc,LIGOScientific:2018gmd,DES:2019ccw,Gray_gwcosmo,O2-H0,GW190814,DES:2020nay,Finke:2021aom,LVK_H0,Palmese_2023}. These events, for which no counterpart is observed, are now called `dark sirens'. Over the preceding years, it has become clear that the source-frame mass distribution of the compact binaries as well as the redshift prior from a galaxy catalogue play a role in a dark standard siren measurement \citep{Taylor2012,Farr:2019twy,Mastrogiovanni:2021wsd}.

Inferring $H_0$ from GW observations may help to resolve the Hubble tension (see e.g. \citealt{Dainotti2021, Dainotti2022}) -- the statistically significant difference between $H_0$ values obtained by methods based on early-Universe measurements, using the cosmic microwave background radiation ($H_0=67.36\pm 0.54\ \mathrm{km}\,\mathrm{s}^{-1}\mathrm{Mpc}^{-1}$; \citealt{Planck_cosmology}) and those obtained from late-Universe measurements, relying on the cosmic distance ladder ($H_0=73.04\pm1.04\ \mathrm{km}\,\mathrm{s}^{-1}\mathrm{Mpc}^{-1}$; \citealt{Cosmology_SN_Riess_2022}). The current best GW measurement of $H_0$ using 47 events from GWTC-3 yields $H_0 = 68^{+12}_{-8}\ \mathrm{km}\,\mathrm{s}^{-1}\mathrm{Mpc}^{-1}$ \citep{LVK_H0}. Given the large uncertainty, this is in agreement with both early-type and late-type methods. In the coming years, an increase in precision of the instruments leading to a higher number of events, as well as improved methodology, will lead to a more precise measurement. As we move towards that goal, 
we need to carefully investigate and mitigate the various systematic effects, both on the GW and on the EM side, which enter the data and the associated assumptions.

It is worth mentioning that standard sirens potentially probe not only the local expansion rate of the universe but also the non-linear Hubble parameter $h(z)$, potentially constraining other parameters of the $\Lambda$-CDM cosmological model such as the matter density fraction $\Omega_m$ or the dark energy equation-of-state parameter $w_0$. In this paper, we only consider the current set of detections. With a handful of nearby observations that we have at the moment, constraints on the latter parameters are not very meaningful (see, {\em e.g.}, Figure~4 of \citealt{LVK_H0}).

It has already been demonstrated in \cite{LVK_H0} that the unknown population distribution of GW sources is the leading source of systematic uncertainty, at least for the GWTC-3 measurement of $H_0$ with the galaxy catalogue method, where the uncertainty in the population has not yet been marginalised over. This is largely because of the relatively low level of completeness of the associated galaxy catalogues -- we are currently dominated by the so-called out-of-catalogue part in the result. The situation may change once we are able to use deeper surveys and the in-catalogue part becomes more important. Potential uncertainties coming from the EM sector have not been extensively studied for dark standard siren measurements, an exception being the work of \cite{DES:2020nay} where two approaches of modelling redshift uncertainties were discussed. It is worth noting that there are several studies on potential systematic effects from peculiar velocities of galaxies \citep[e.g.,][]{peculiar_vel_Howlett, pec_vel_Mukherjee, peculiar_vel_Nicolaou}. However these effects are relevant particularly for a $H_0$ measurement coming from a handful of nearby events (i.e., bright standard sirens). Peculiar velocities are relatively small for bright standard sirens which are typically more distant events, and they are further expected to average or cancel out over a large number of observations required for a reasonable $H_0$ measurement. For dark standard sirens, redshift measurement uncertainties are thus expected to be the significant source of error in the EM sector.

In current dark siren measurements of $H_0$, the galaxy catalogues used include only a small fraction of exact spectroscopic redshifts. Most of the redshifts in such wide-angle datasets are instead estimated from galaxy multi-band photometry (photometric redshifts, photo-$z$s). Despite such extensive spectroscopic surveys as DESI \citep{DESI}, 4MOST \citep{4MOST} or Euclid \citep{Euclid}, this situation is unlikely to change in the coming years. 
Photometric redshifts bear considerable uncertainties (typically 10\% or more) and this can influence related analyses.
In this paper, we investigate the impact of redshift uncertainty models on the dark standard siren $H_0$ measurement. We first study redshift uncertainty models which capture more features than a simple Gaussian profile. In particular, we model the uncertainty in the redshift of a potential host galaxy as a modified Lorentzian distribution and fit its parameters to data from the 2 Micron All-Sky Survey Photometric Redshift Catalogue (2MPZ, \citealt{Bilicki_2MPZ}) and the WISExSCOS Photometric Redshift Catalogue (WISC, \citealt{Bilicki_WISC}). Subsequently, we modify the standard \texttt{gwcosmo} pipeline used for $H_0$ inference \citep{Gray_gwcosmo,Gray_pixelated} to include different redshift uncertainty models other than the previously-used simplistic Gaussian model. Finally, in order to estimate the impact of redshift uncertainties, we compute $H_0$ using the 47 events from GWTC-3 used in \cite{LVK_H0} and the different uncertainty models for each of the two catalogues above. 

The rest of this paper is organised as follows. We describe our choice of galaxy catalogues in Section~\ref{sec:catalogs} and go over our modelling of redshift uncertainties in Section~\ref{sec:redshift}. We review our $H_0$ inference method and present the settings and the associated GW data in Section~\ref{sec:method}. We describe and discuss our results in Section~\ref{sec:results}. In the same section, we also include a brief discussion of how redshift uncertainties can impact the in-catalogue and out-of-catalogue components of a $H_0$ inference method. We conclude and mention a few items for follow-up work in Section~\ref{sec:conclusion}. There are two appendices which provide additional details on some specific aspects of Section~\ref{sec:redshift} and Section~\ref{sec:results} respectively.

\section{Electromagnetic data} \label{sec:catalogs}

The GLADE+ galaxy catalogue \citep{Dalya_glade+} was used to obtain electromagnetic 
data for the most recent GW $H_0$ inference by the LVK collaboration. GLADE+ is an expanded version of the GLADE sample 
\citep{Dalya_Glade}, which had previously been used in several cosmological studies (see e.g.~\citealt{Fishbach_H0,GW190814,O2-H0}). As GLADE+ is a composite galaxy catalogue created by combining data from six different astronomical databases \citep{GWGC,Bilicki_2MPZ,2MASS,HyperLEDA,Bilicki_WISC,SDSS_DR16_Q} with different properties, such a sample could not be easily applied within our methodology, which requires a particular photo-$z$ error model for a given galaxy dataset (see Section~\ref{sec:redshift}). 
Hence, here we use two of the major constituent catalogues of GLADE+, 2MPZ and WISC. 
These two catalogues offer the main advantages of covering very large areas of the sky and having well-controlled selections and photo-$z$ performance; this is especially the case for 2MPZ. They also have a number of drawbacks, though, such as optical photometry from the photographic era, very limited depth (especially in 2MPZ), and occasionally complicated source selection, joining the optical, near-, and mid-infrared. These issues are now very much resolved in new datasets such as Pan-STARRS \citep{Pan-STARRS} or the DESI Legacy Imaging Survey \citep{DESI_LS} and we plan to use such updated samples for related studies in the future. For this work, however, the 2MPZ and WISC catalogues will suffice as a good test-bed for leading order effects of photo-$z$ uncertainties on $H_0$ derivation from GW events within the statistical method.

Both 2MPZ and WISC are multi-band photometric catalogues that were created by combining all-sky legacy optical photometry obtained from digitised photographic plates (SuperCOSMOS; \citealt{SCOS,SCOS_2}) with infrared datasets. In the 2MPZ case, these latter were the Two Micron All Sky Survey Extended Source Catalogue (2MASS XSC; \citealt{2MASS,2MASS_2}) and the Wide-field Infrared Survey Explorer (WISE; \citealt{WISE}) from its "All-Sky" data release \citep{WISE_2}. This provided nearly 1 million galaxies over most of the sky with 9-band photometric coverage, including photographic $B_J, R, \mathrm{and}\ I$, near-IR $J, H, \mathrm{and}\ K_s$, and mid-IR $W1$ and $W2$. In WISC, 2MASS was no longer used, while WISE data originated from the subsequent "AllWISE" catalogue \citep{WISE_3}. The resulting cross-match includes nearly 20 million galaxies, reliably covering about 70\% of the sky in four bands, $B_J$ $R$ $W1$ $W2$. In the 2MPZ case, its depth is driven by the shallowest of the 3 input datasets, 2MASS XSC, and it is effectively a $K_s$-band-selected sample, flux-limited to $K_s<13.9$ mag (Vega). Its median redshift is $z=0.07$ and it has practically no galaxies beyond $z\sim0.25$. WISC has a slightly more complicated selection, being flux-limited in three bands jointly: $B_J < 21$, $R < 19.5$ (both AB-like), and $W1 < 17$ (Vega). It is about 3 times deeper than 2MPZ, with $z_\mathrm{med}\simeq0.2$ and reaching up to $z\sim0.45$.

The multi-band photometry of 2MPZ and WISC was further employed to obtain photo-$z$s for all the included galaxies. This was done via supervised machine-learning (ML) methodology, where a model is "trained" on ground-truth data to deliver predictions for a given quantity in an output sample. In this case, the mapping of interest is between photometry and redshift, and the ML method used was artificial neural networks (ANN) as implemented in the public "ANNz" package \citep{ANN}. The training sets include galaxies that overlap between the photometric datasets and the external spectroscopic redshift (spec-$z$) samples; the latter redshift measurements are typically accurate enough to assume that they have no errors that could influence the ML training. The spec-$z$ data used for the training sets was chosen in a way to best match the depth of the photometric samples. For 2MPZ, these were mostly based on the 2MASS Redshift Survey \citep{2MRS}, the 6dF Galaxy Survey \citep{6dF}, the 2dF Galaxy Redshift Survey \citep{2dF} and the Sloan Digital Sky Survey Data Release 9 \citep{SDSS}. In WISC, the photo-$z$ training sample was derived from a cross-match with the Galaxy And Mass Assembly \citep{GAMA} from its "GAMA-II" catalogues \citep{GAMA_2}. Additionally, SDSS DR12 \citep{SDSS_2} was used to calibrate the purification of the WISC dataset from stars and quasars.

\section{Modeling the redshift uncertainties} \label{sec:redshift}
Photometric redshifts often have large uncertainties and non-trivial error profiles. Using an inaccurate uncertainty profile may lead to a bias in $H_0$~\cite{DES:2020nay}, so reliable photo-$z$ error estimates are needed. In an ideal scenario, one could use uncertainty quantification provided for instance by the photo-$z$ algorithm,  for each of the considered EM sources. This was for instance applied in \cite{DES:2020nay} for the Dark Energy Survey (DES) galaxies and the associated DNF photo-$z$ estimation algorithm. There the authors compare the case when individual photo-$z$ probability density functions are used, to the case of Gaussian approximation. Out of the two events they consider, only one leads to a notable difference in the $H_0$ posterior. Hence, it still remains to be seen how important this approach would be for a considerably larger dataset. Unfortunately, most of the photo-$z$ methodologies, especially those ML-based, do not reliably deliver such information for the individual objects. This is not different in the case of ANNz-derived photo-$z$s in the 2MPZ and WISC catalogues which we consider. What can be however done, is to calibrate the photo-$z$ uncertainties \textit{a posteriori}, by comparing the photo-$z$s with overlapping spec-$z$ samples and analyzing the redshift residuals, $\Delta z \equiv z_\mathrm{photo} - z_\mathrm{spec}$ (often additional rescaled by $1+z$). In practice, such a model cannot usually be obtained for each individual source separately. It is however possible to build more general photo-$z$ error models as a function of redshift, magnitude or possibly also galaxy color. In this work, we look at this first option and construct photo-$z$ error models for 2MPZ and WISC as a function of photometric redshift, which is the most relevant as an `observable' (unlike spec-$z$, which is not available for most galaxies in a photo-$z$ dataset). 
These photo-$z$ error models provide the posterior distribution on the true redshift of each galaxy given the measured $z_\text{photo}$, $p(z|z_\text{photo})$.

\begin{figure}
    \centering
    \includegraphics[width=0.5\textwidth]{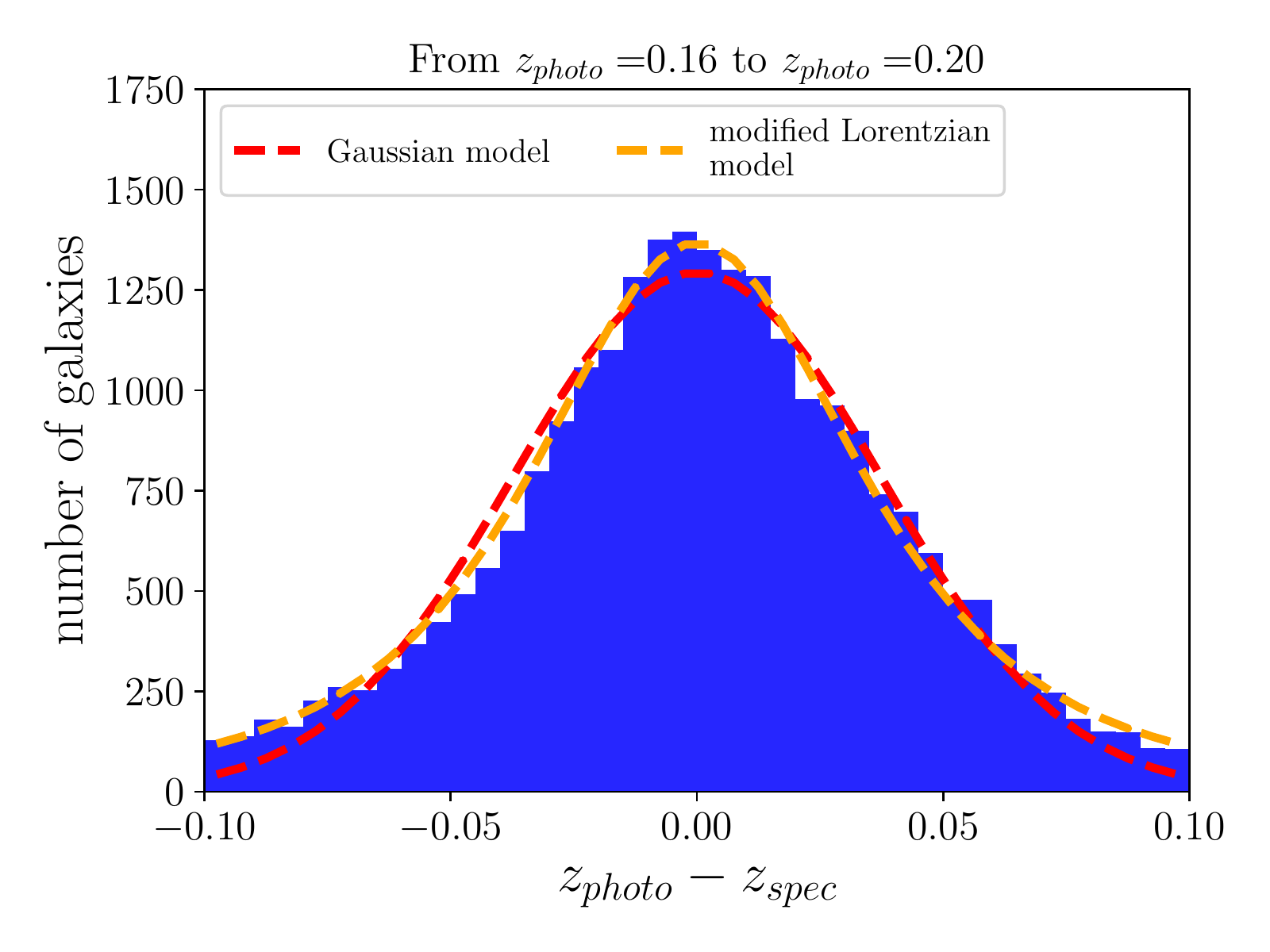}
    \caption{Comparison between Gaussian and modified Lorentzian model of uncertainties. Blue histogram shows the difference between photo-$z$ and spec-$z$ for WISC catalogue for galaxies between $z_\mathrm{photo} > 0.28$ and $z_\mathrm{photo} < 0.32$. Red (orange) line is a Gaussian (modified Lorentzian) model fitted to the histogram.}
    \label{fig:gaussvslorentz}
\end{figure}

The simplest photo-$z$ error model is a Gaussian, with some mean residual $\mu$ and scatter $\sigma$. It is however well known from the literature that even `well-behaved' photo-$z$s (i.e. those for which errors are symmetric and centered at 0), usually have non-Gaussian tails. For instance, for the 2MPZ case, a better fit is obtained if a `modified Lorentzian' is used \citep{Bilicki_2MPZ}.  
This distribution is given by
\begin{equation}
    f(\Delta z)=A\left( 1+\frac{\Delta z^2}{2as^2} \right)^{-a}
    \label{eq:mlorentz}
\end{equation}
where $\Delta z$ is a difference between the photometric and spectroscopic redshift, $A$ is a normalization factor, while $a$ and $s$ are free parameters that need to be inferred for the galaxy catalogue. Note that this is valid only when $a>0.5$. The same model, albeit with different parameters than for 2MPZ, was also used for modelling the WISC photo-$z$ uncertainties in \cite{2018MNRAS.481.1133P}. The difference between Gaussian and modified Lorentzian uncertainty model is shown in Figure \ref{fig:gaussvslorentz}. 
In such a formulation, the average photo-$z$ errors are assumed to be 0, which for 2MPZ and WISC is a sufficiently good approximation in the sense that the mean bias is much smaller than the scatter. We note however that the model could be generalised for non-zero average bias (e.g. \citealt{2021MNRAS.501.1481H}).
Finally, other analytical formulations could be used to recover the non-Gaussian features, such as for instance the Student-$t$ distribution (e.g. \citealt{2020arXiv200813154V}) or other empirically-driven fits \citep{2018MNRAS.476.1050B}.

As mentioned above, we will be modelling the photo-$z$ uncertainties as functions of the `'observed' photo-$z$ itself. This will be done both for the simpler Gaussian case, where we will fit for both $\mu$ and $\sigma$ (that is, our Gaussian models do not have to be centred on 0), and for the modified Lorentzian in the unbiased case (i.e. with free parameters $a$ and $s$). In the most general case, all these parameters can be some general functions of photo-$z$, although we will limit ourselves to their linear evolution with $z_\mathrm{photo}$. For that, we use spectroscopic calibration samples, thanks to which we can calculate redshift residuals $\Delta z$ for the full range of $z_\mathrm{photo}$ of a given dataset. For the methodology to work properly, these calibration samples should be representative for the whole photometric dataset in question. This would ideally mean a random subsample of the entire catalogue, which is hardly ever available. The second-best scenario, that our calibration samples meet, is to use spec-$z$ data from a flux-limited sample deeper than the calibrated photo-$z$ catalogue. For 2MPZ, we employ all the spec-$z$ samples overlapping with it, mentioned in the previous Section, the most useful of them being SDSS, which is deeper than 2MASS XSC. For WISC, such a calibration sample is provided by GAMA, which in its equatorial fields is a very complete, flux-limited (98\% at $r<19.6$, \citealt{2022MNRAS.513..439D}) galaxy sample.

To build the photo-$z$ error models, we divide the joint spec-$z$ -- photo-$z$ data into photo-$z$ bins with a width of $\Delta z=0.02$ for 2MPZ and $\Delta z=0.04$ for WISC
and fit both Gaussian and modified Lorentzian profiles to $\Delta z$ distributions obtained for the particular bins to infer parameters $a$, $s$, $\sigma$ and $\mu$ for every bin. Then we fit a linear function for each parameter in $z_\mathrm{photo}$. 
Results of our calibration are provided in Appendix~\ref{app:model}, and are shown for modified-Lorentzian and Gaussian models for the 2MPZ and WISC catalogues respectively in Figures \ref{fig:2MPZ_params} and \ref{fig:WISC_params}. We set $\mu=0$ for the rest of this paper for both catalogues because the fitted value is small relative to redshift. Then we refit $\sigma$ with that assumption. The obtained model is shown in Table \ref{tab:params}. 
We apply the attained model to all galaxies in the catalogues. Obtained distributions are truncated at $z = 0$ and rescaled because 
redshift below zero is not physical. Moreover, the modified Lorentzian equation is only valid if $a>0.5$ which corresponds to redshift $z_\mathrm{photo} < 0.34$ for 2MPZ and $z_\mathrm{photo} < 0.424$ for WISC. There are not many galaxies outside of that range so we put a redshift cut there. 

\begin{table}
\label{tab:params}
\begin{tabular}{l|ll}
         & \multicolumn{1}{c}{2MPZ}        & \multicolumn{1}{c}{WISC}        \\ \hline
$\sigma$ & $0.052z_{\mathrm{photo}}+0.008$ & $0.085z_{\mathrm{photo}}+0.019$ \\
$a$      & $-10.1z_{\mathrm{photo}}+3.9$   & $-4.74z_{\mathrm{photo}}+2.51$  \\
$s$      & $0.031z_{\mathrm{photo}}+0.01$  & $0.043z_{\mathrm{photo}}+0.021$
\end{tabular}
\caption{The obtained relation of parameters $\sigma$, $a$, $s$ and photometric redshift $z_{\mathrm{photo}}$ for 2MPZ and WISC catalogues.}
\end{table}


This way we have a model for photo-$z$ uncertainty for each galaxy included in a given catalogue. We note however that in the future this approach may need to be more sophisticated, for instance by building different photo-$z$ models for various galaxy populations. For instance, red galaxies are known to have generally better photo-$z$ performance than the blue ones (e.g. \citet{Bilicki_KiDS}) and such details may need to be accounted for as a secondary effects as the statistical error in the $H_0$ measurement shrinks with an increasing number of GW events.

\section{Cosmology inference} \label{sec:method}

\subsection{Methodology and codebase}

We employ the pixelated version of the \texttt{gwcosmo} code \citep{Gray_thesis,Gray_pixelated} to calculate $H_0$ from GW data using the galaxy catalogue method. The essential methodology behind the inference is described in \cite{Gray_gwcosmo,O2-H0}. The posterior probability density on $H_0$ is expressed as a prior $p(H_0)$ times the likelihood from each of the individual events $\text{ev}$, as
\begin{align}
p(H_0|\{x_{\text{GW}}\},\{D_{\text{GW}}\}) \propto p(H_0)p(N_{\text{det}}|H_0)\prod_{\text{ev}=1}^{N_{\text{det}}} p(x_\text{GW}^\text{ev}|D_\text{GW}^\text{ev},H_0)\,.
\label{eq:posterior}
\end{align}
Here $D_\text{GW}^\text{ev}$ (subsequently $D_\text{GW}$) indicates a positive GW detection and $x_\text{GW}^\text{ev}$ (subsequently $x_\text{GW}$) is the associated GW data. The additional term $p(N_{\text{det}}|H_0)$ which appears in front of the expression is the probability of detecting $N_\text{det}$ events; it is marginalized over assuming a uniform in log rate prior $p(R)\propto R^{-1}$, and loses it dependence on $H_0$ \citep[for more details, see][]{Fishbach:2018edt,O2-H0}.

For each event, the likelihood separates into in-galaxy-catalogue and out-of-galaxy-catalogue parts (denoted respectively by $G$ and $\bar{G}$ below). What is contained in the galaxy catalogue and what is not is governed by the depth of the galaxy survey(s). To the first order, this corresponds to an apparent magnitude threshold $m_\text{th}$, which can potentially vary across the sky. The pixelated version of \texttt{gwcosmo} allows for such a variation of $m_\text{th}$. The sky is first divided into $N_\text{pix}$ equally-sized pixels following \texttt{HEALPix} \citep{HEALPix}, and the likelihood is computed by summing over the contributions for each of these pixels $i$.
\begin{align}
p(x_\text{GW}|D_\text{GW},H_0) = & \frac{1}{N_\text{pix}} \sum_{i=1}^{N_\text{pix}} {\Bigl[} p_i(x_\text{GW}|G,D_\text{GW},H_0)p_i(G|D_\text{GW},H_0) \nonumber \\ & + p_i(x_\text{GW}|\bar{G},D_\text{GW},H_0)p_i(\bar{G}|D_\text{GW},H_0) { \Bigr]}  \,. 
\label{eq:in-out}
\end{align}
Within pixel~$i$, $p_i(G|D_\text{GW},H_0)$ is the probability that the host galaxy is in the catalogue and $p_i(\bar{G}|D_\text{GW},H_0)\equiv1-p_i(G|D_\text{GW},H_0)$ the complementary probability that the host galaxy is not in the catalogue.

The in-catalogue likelihood is evaluated as a sum over the redshift distribution of each of the individual galaxies in the catalogue, times the support of the GW likelihood in that sky-direction and distance (given $H_0$) times the probability of the given galaxy to host or to {\em source} the GW event:

\begin{align}
& p_i(x_\text{GW}|G,D_\text{GW},s,H_0) = \nonumber \\ &\frac{ \displaystyle \sum_{j=1}^{N_\text{gal}}\int dz \, p(x_\text{GW}|z,\Omega_j,s,H_0) \,p(z|z_\text{photo}^j) \,p(s|z)\,p(s|M(z,m_j,H_0))}{\displaystyle \sum_{j=1}^{N_\text{gal}}\int dz \, p(D_\text{GW}|z,\Omega_j,s,H_0) \,p(z|z_\text{photo}^j)  \,p(s|z)\,p(s|M(z,m_j,H_0))}\,.
\label{eq:in-catalog-sum}
\end{align}

The parameter $s$ on the right-hand side of the initial expression is to state the (previously implicit) assumption that there is a real GW source associated with the GW data. Here $p(z_j)$ is the probability distribution of the redshift of the $j$\textsuperscript{th} galaxy in the catalogue, and $\Omega_j$ its sky-direction. The standard \texttt{gwcosmo} code assumes a Gaussian for the redshift distribution of the galaxy. We modify the previous codebase so that it is able to include more generic redshift distributions. The second term within the integral above can be obtained from standard GW parameter inference which yields $p(x_\text{GW}|d_L,\Omega)$ and using $d_L=d_L(z,H_0)$, with $d_L$ being the luminosity distance. The third term within the integral is the probability of a galaxy at a redshift $z$ to host a GW source $p(s|z)$, evaluated at the galaxy redshift $z=z_j$. This probability is the GW event rate in the source frame $R(z)$ multiplied by the cosmological time dilation factor for conversion from the source frame to the detector frame, $p(s|z)  \propto (1+z)^{-1}R(z)$.
The final term within the integral, $p(s|M)$, is the probability of a galaxy of absolute magnitude $M$ to host a GW event. This can be used to weight the galaxies in proportion to their luminosities (in a certain observed band),
\begin{align}
p(s|M(z_j,m_j,H_0)) \propto L(M_j(H_0))\,.
\label{eq:lum-weight}
\end{align}
Here $m_j$ is the apparent magnitude of the galaxy $z_j$ and $M_j(H_0)\equiv M(z_j,m_j,H_0)$ is given by the standard expression
\begin{align}
M_j(H_0)=m_j-5\log_{10}\left(\frac{d_L(z_j,H_0)}{\text{Mpc}}\right)-25\,.
\label{eq:dist-modulus}
\end{align}

The denominator outside the overall expression is the GW selection function or the {\em detection efficiency} \citep{GW170817_H0,Mandel:2018mve} which quantifies the probability of a GW detection (given $H_0$), and is obtained formally by integrating over all detectable GW data sets $\{x_\text{GW}\}$:
\begin{align}
p_i(D_\text{GW}|G,s,H_0) = \int dx_\text{GW} p_i(x_\text{GW}|G,s,H_0)\,.
\label{eq:det-eff}
\end{align}
In practice, the computation of a selection function involves assuming priors on the source distribution (including mass and redshift distributions) and integrating over the cases where the GW event is above a certain signal-to-noise ratio (SNR) threshold and can be assumed to be {\em detected}.

The out-of-catalogue likelihood is evaluated as an integral instead of a sum over discrete galaxy redshifts, assuming a redshift and sky distribution of the unobserved galaxies and a magnitude distribution of their luminosities, in proportion to their probability of hosting a GW source:

\begin{align}
&p_i(x_\text{GW}|\bar{G},D_\text{GW},s,H_0) = \nonumber\\& \frac{\int dz \, d\Omega \, dM p(x_\text{GW}|z,\Omega,s,H_0)\, p(z)\,p(\Omega)\,p(M|H_0) \, p(s|z)\,p(s|M)}{\int dz \, d\Omega \, dM p(D_\text{GW}|z,\Omega,s,H_0)\, p(z)\,p(\Omega)\,p(M|H_0) \, p(s|z)\,p(s|M)}
\,.
\label{eq:out-catalog-integral}
\end{align}

The prior on the redshift is taken to be proportional to the differential comoving volume, $p(z)\propto dV_c(z)/dz$.
The sky location prior $p(\Omega)$ is taken to be uniform in the sky. The prior on the absolute magnitude of the galaxies is chosen to be a Schechter function
\begin{align}
p(M|H_0)\propto 10^{-0.4(\alpha+1)(M-M^*(H_0))}\exp\left[{-10^{-0.4(M-M^*(H_0))}}\right]\,,
\label{eq:schechter}
\end{align}
described by a characteristic magnitude $M^*$ and a slope $\alpha$ for the given observation band. The remaining terms are obtained as above.

Once again, the denominator obtained by integrating over all the ``gravitationally'' detectable datasets,
\begin{align}
p_i(D_\text{GW}|\bar{G},s,H_0) = \int dx_\text{GW} p_i(x_\text{GW}|\bar{G},s,H_0)\,.
\label{eq:det-eff-bar}
\end{align}

To complete the discussion of the methodology, the in-catalogue probability for each of the pixels is obtained using
\begin{align}
& p_i(G|D_{\text{GW}}, s,H_0)
= \dfrac{\iiint^{z(m_{\text{th}},M,H_0)}_0 
dz\,d\Omega\,dM\,I(z,\Omega,M)}
{\iiint_0^\infty dz\,d\Omega\,dM\,I(z,\Omega,M)},\ \text{with} \nonumber \\
& I(z,\Omega,M) \equiv p(D_{\text{GW}}|z,\Omega, M, H_0) p(z)p(\Omega)p(M|H_0 ) p(s|z) p(s|M)\,.
\label{eq:pG}
\end{align}

\subsection{Data and parameter settings} \label{sec:settings}

We stick to the settings of \citet{LVK_H0} wherever possible. In particular, we choose the same set of GW events with a signal-to-noise ratio (SNR) $> 11$. However, since we are interested in redshift uncertainties coming from galaxy catalogues, we consider only the ``dark'' standard sirens; in particular, we do not include GW170817 in our analysis. As a result, we have 46 events: 42 binary black holes, together with the asymmetric mass binary GW190814, neutron star black hole binaries GW200105 and GW200115, and binary neutron star GW190425. 

For each of these events, we use the GW selection function as in \citet{LVK_H0}. The computation of a selection function requires an assumption on the mass distribution of the binaries, their rate evolution, and the sensitivity of the detector network. For the mass distribution we use a power law + peak model \citet{LVK_O3b_astrodist} with  the power law slope of the primary mass distribution $\alpha=3.78$, the power law slope of the secondary mass distribution $\beta=0.81$,  the primary mass in the mass range
$m_\text{max} = 112.5 M_\odot$, $m_\text{min} = 4.98 M_\odot$,  with a window scale at the lower mass end $\delta_m = 4.8 M_\odot$, mean of the Gaussian peak $\mu_g = 32.27 M_\odot$, standard deviation of the Gaussian peak $\sigma_g = 3.88 M_\odot$ and the relative weight of the Gaussian peak with respect to the power law quantified by the parameter $\lambda_g = 0.03$. For the rate evolution, we use a Madau-Dickinson model \citep{Madau:2014bja} with a low-redshift power-law slope $\gamma = 4.59$, a high-redshift power-law slope $k = 2.86$ and a peak at $z_p = 2.47$  separating the two regimes. These are the median values for the joint population-cosmology analysis (without galaxy catalogues) performed in \cite{LVK_H0} using \texttt{icarogw} \cite{Mastrogiovanni:2021wsd}. We further use the sensitivities from the O1, O2, and O3 observing runs of LIGO and Virgo. The same source-frame prior distributions are also used to reweight the GW posterior samples for each of the 46 events considered.

It is worth noting that changing the mass distribution and the rate evolution parameters from their central values can lead to significant differences in the final result. This was studied in \cite{LVK_H0}, and the results were reported in Figure 11 there. In the meantime, approaches have been developed to {\em marginalize over} a distribution of these parameters while simultaneously estimating $H_0$ \citep{Mastrogiovanni:2023emh,Mastrogiovanni:2023zbw,Gray:2023wgj}. The assumption of fixed values of mass and rate parameters for an $H_0$ inference will thus no longer be necessary.

We make a different choice for the galaxy catalogue however, namely, we use 2MPZ and WISC instead of GLADE+. We use the $B_J$ band with Schechter function parameters $M^*=-19.66$ and $\alpha=-1.21$ from \citet{schechter_params} for both 
catalogues.

In the first set of runs, we use the standard \texttt{gwcosmo} code with Gaussian uncertainties. In the next set of runs, we change the redshift uncertainty profile from a Gaussian to the modified Lorentzian, the rest of the setup being the same as previously. For the third set of runs, we turn off the uncertainties on the redshift altogether. Finally, we artificially alter the redshift uncertainties. For Gaussian uncertainties, this corresponds to multiplying the standard deviation $\sigma$ by a constant. For the modified Lorentzian case boosting the uncertainties is not straightforward, as the parameters $a$ and $s$ are related, and moreover, the model is only valid if $a>0.5$, so we artificially shrink the uncertainties by using $a(\frac{1}{2}z_{\mathrm{photo}})$ and $s(\frac{1}{2}z_{\mathrm{photo}})$, for every galaxy with redshift $z_{\mathrm{photo}}$.

\section{Results and Discussion} \label{sec:results}

\begin{figure}
    \centering
    \includegraphics[width=0.5\textwidth]{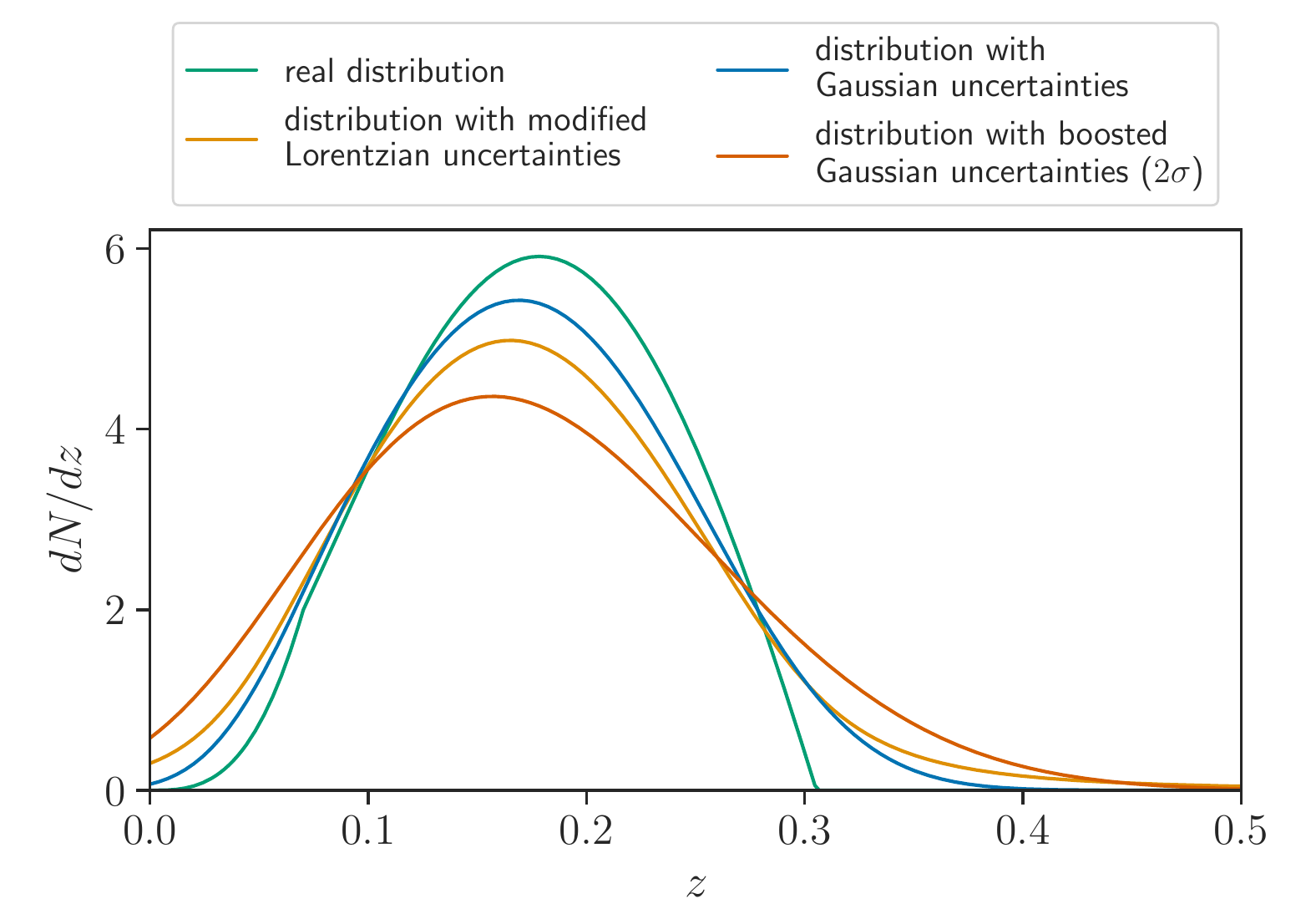}
    \caption{The distribution of observed redshifts in a mock catalogue mimicking the WISC magnitude limit of $m_{B_J}=21$ mag.
    We use a background cosmology and model the luminosity distribution of the galaxies as a Schechter function with parameters discussed in the text, and generate the `true' redshift distribution shown with the green line. Then we perturb it with Gaussian (blue line) and modified Lorentzian (yellow line) uncertainties as in the real WISC catalog.}
    \label{fig:mockcat}
\end{figure}

In this Section we present the result of applying the photo-$z$ error models described in Sec.~\ref{sec:redshift} to the $H_0$ derivation framework detailed in Sec.~\ref{sec:method}, based on the 2MPZ and WISC galaxy catalogs. We start however by discussing how the redshifts of different quality influence the observed distribution of galaxies.  For that, we generate a mock catalog of uniformly distributed galaxies in the flat $\Lambda$CDM background cosmology with $H_0=70\mathrm{km}\,\mathrm{s}^{-1}\mathrm{Mpc}^{-1}$ and $\Omega_m=0.3$. We assume that galaxies follow a \citet{schechter1976} luminosity function with same parameters as mentioned in Sec. \ref{sec:settings}.
Then we assume that the catalogue is magnitude limited with a threshold $m^{B_J}_\text{obs}=21$ mag. This gives us the `true' redshift distribution, shown with the green line in Fig.~\ref{fig:mockcat}. Then we perturb these `true' redshifts using the  Gaussian and modified Lorentzian redshift uncertainties as calculated in Sec.~\ref{sec:redshift} for the WISC catalog. Applying the uncertainties to a catalogue shifts the whole redshift distribution, as well as its peak, to lower values, and also adds a tail at high redshifts. This shift may influence the statistical derivation of $H_0$ along with other analysis in which redshift statistics are involved. 

\begin{figure}
    \centering
    \includegraphics[width=0.5\textwidth]{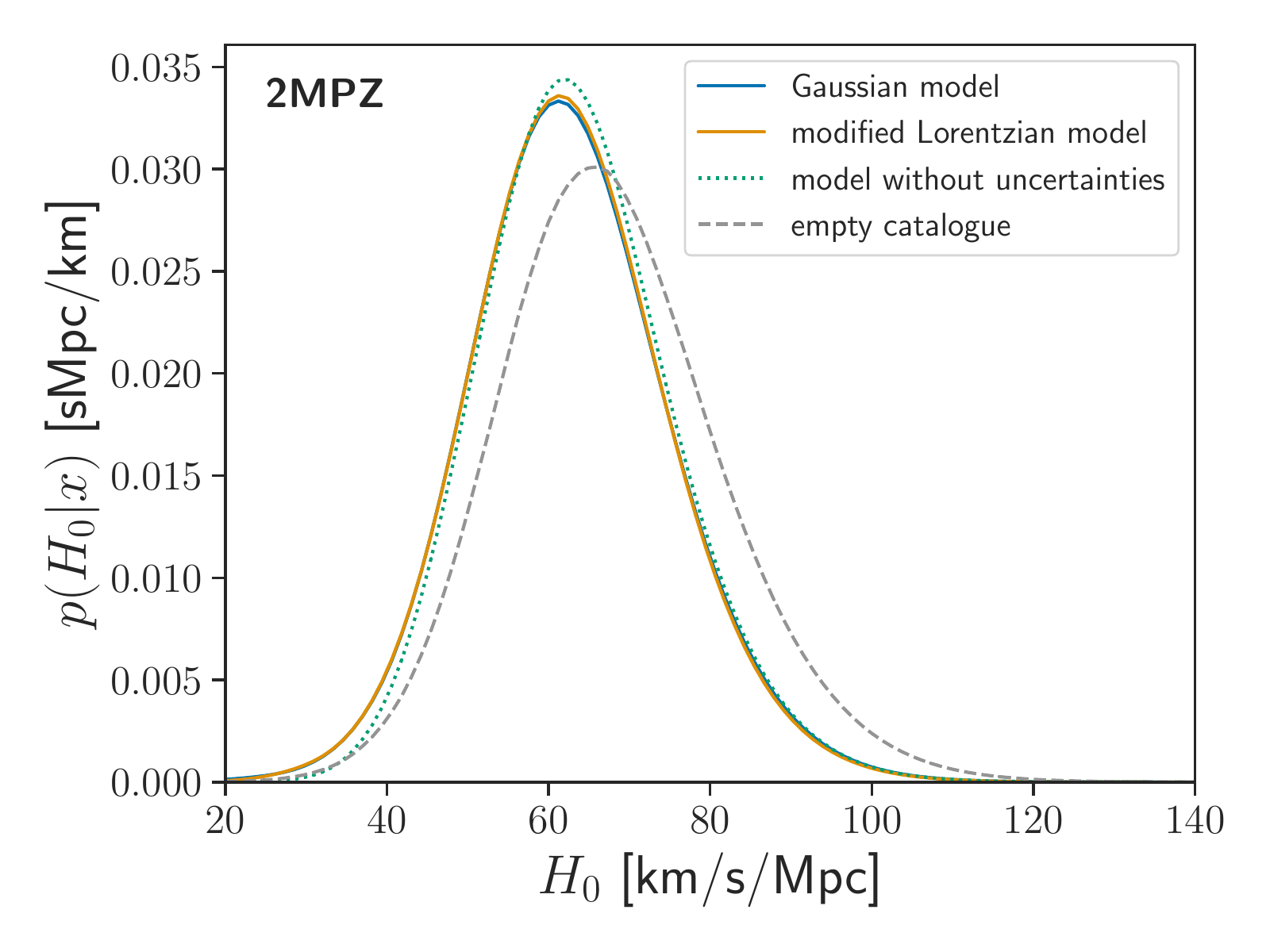}
    \caption{The influence of different redshift uncertainty models on the inference of the Hubble constant based on the 2MPZ catalog. We plot the posterior probability distribution in $H_0$ obtained using the 46 events used in \protect\cite{LVK_H0}. Results with the Gaussian and modified Lorentzian uncertainty models are plotted in blue and orange, respectively. For reference, we also show the result without any uncertainty on the redshift (dotted green line). As well as an empty catalogue case (dashed gray line).}
    \label{fig:2MPZresult}
\end{figure}

To study the influence of the photo-$z$ error models on the $H_0$ posterior we start by the shallower 2MPZ catalog. Although it covers almost the entire sky, and has better-quality redshifts than WISC, it is very shallow when compared to the typical distances of GW events detected by LVK. Only 19 events out of 46 we use from GWTC-3 have estimated distances within 2MPZ coverage of $z<0.25$ ($d_L\simeq 1$ Gpc). Therefore, $H_0$ derivations using this catalog will be dominated by the out-of-catalog terms and in particular the population model. Indeed, as shown in Fig.~\ref{fig:2MPZresult}, the impact of various photo-$z$ models is minimal on the $H_0$ posterior in that case. The modified Lorentzian and Gaussian case give practically the same results, and the idealized case of no uncertainties is only slightly different. This latter is not representing the actual situation, as it assumes that the redshifts reported in the 2MPZ catalog are exact. We show it just for comparison as an `upper limit', equivalent to the hypothetical case of spectroscopic redshifts with negligible uncertainties. We also add the empty catalogue case: a `lower limit', with a redshift prior coming only from the background cosmology, and with no information from galaxies. The main results of this comparison is that catalogs of such a depth as 2MPZ  do not provide enough information to significantly change the posterior with that method, therefore we do not further analyse this case.

\begin{figure}
    \centering
    \includegraphics[width=0.5\textwidth]{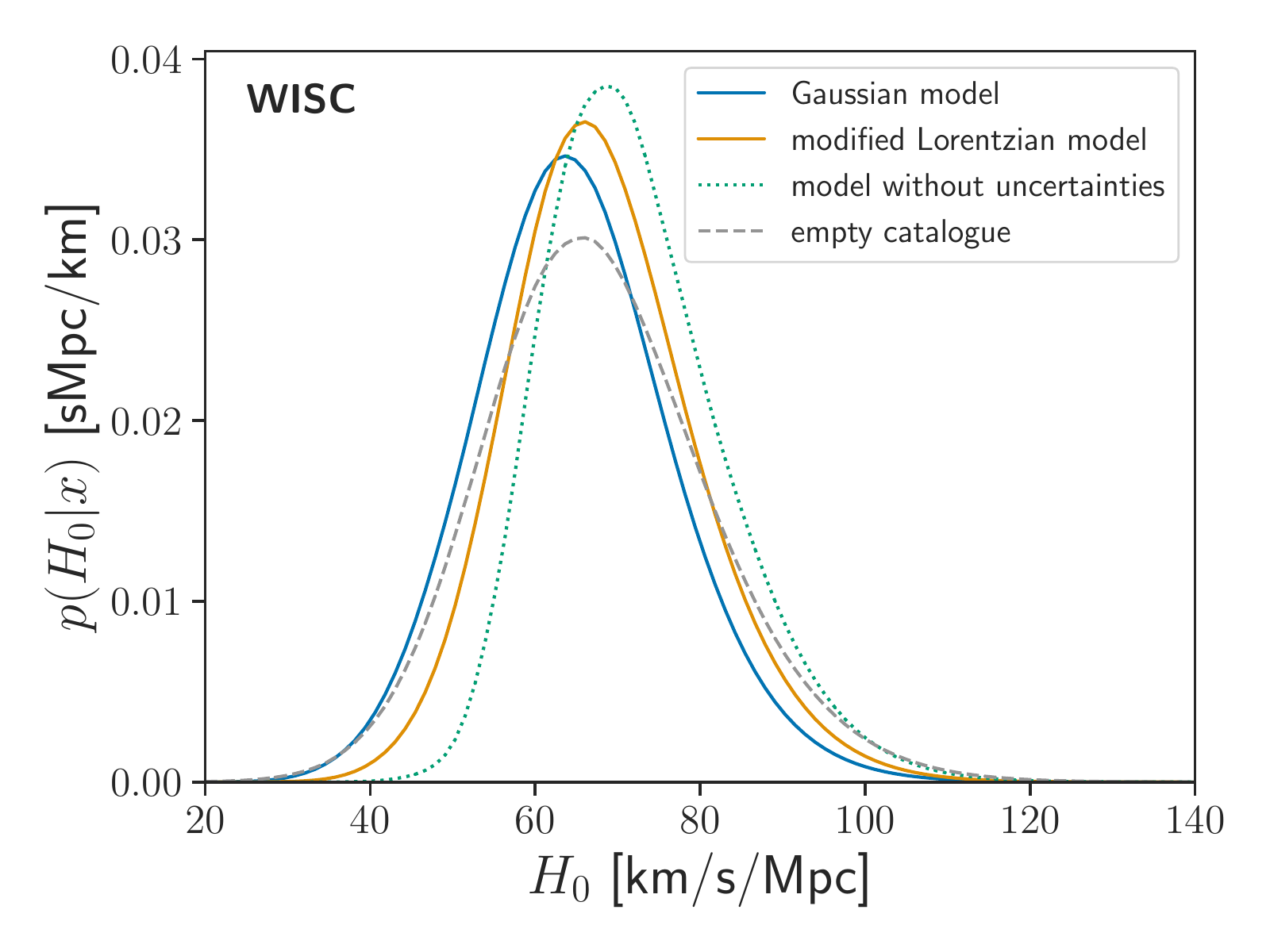}
    \caption{The influence of different redshift uncertainty models on the inference of the Hubble constant based on the WISC catalog. We plot the posterior probability distribution in $H_0$ obtained using the 46 events used in \protect\cite{LVK_H0}. Results with the Gaussian and modified Lorentzian uncertainty models are plotted in blue and orange respectively. For reference, we also show the result without any uncertainty on the redshift (dotted green line), as well as for the empty catalog case (dashed grey).}
    \label{fig:WISCresult}
\end{figure}

We now turn to the comparison of different redshift uncertainty models for the WISC catalogue, shown in Figure \ref{fig:WISCresult}. 
Moreover we add already discussed upper-limit case of no redshift uncertainties and lower-limit of empty catalogue.
Introducing Gaussian or modified Lorentzian uncertainty models moves the peak of the distribution to support lower values of $H_0$. This can be understood by recalling Figure \ref{fig:mockcat}. Introducing the uncertainty to the galaxy catalogue impacts the distribution of the observed redshifts.
As detailed in Sec.~\ref{sec:redshift}, the uncertainties of our models grow with redshift, therefore low-redshift galaxies are less impacted by the scatter. However the change in uncertainty is significant at the peak of the redshift distribution by scattering galaxies from that area to lower and higher redshifts. Galaxies from the high end of the redshift range have large uncertainties, thus they can be scattered into very high redshifts causing the tail of the distribution to appear.
All that causes the peak of the distribution to move to lower redshift values and thus supporting lower values of $H_0$. Moreover, there is an additional effect for the modified Lorentzian case. The truncation of the redshift posterior of each galaxy at $z=0$ causes the mean of the posterior to move to higher values, which correspond to supporting higher values of $H_0$. This truncation is significant for modified Lorentzian distribution but not for Gaussian distribution. That is the reason why the modified Lorenzian model peak of $H_0$ is to the right of the Gaussian model peak. 

\begin{figure}
    \centering
    \includegraphics[width=0.5\textwidth]{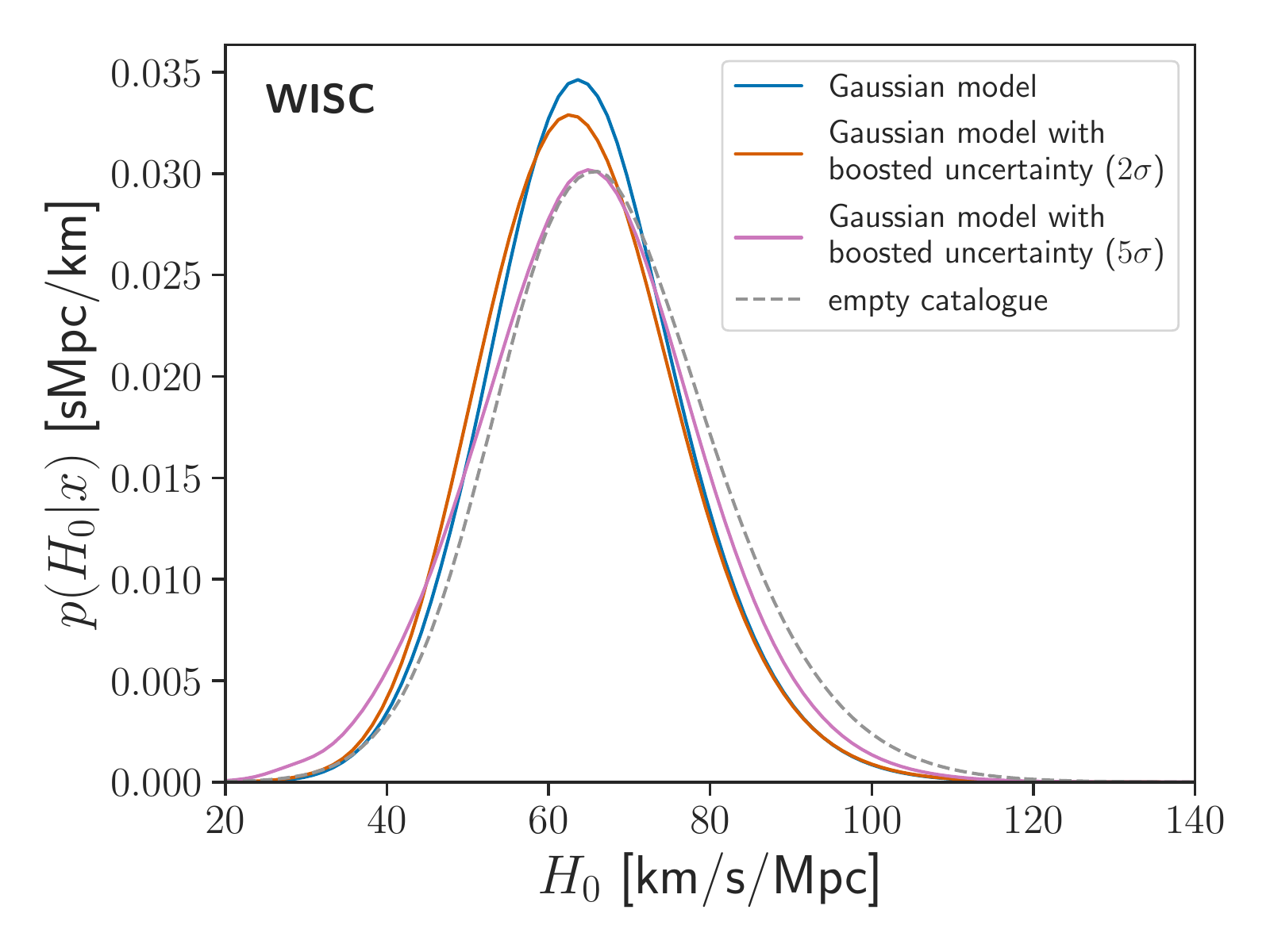}
    \caption{The influence of boosting the redshift uncertainty in the Gaussian model for the WISC catalogue. We plot the $H_0$ posterior density with the Gaussian model as fitted to date (blue), with the Gaussian model but uncertainties enhanced by a factor of 2 (orange), and one with the uncertainties enhanced by a factor of 5 (pink).}
    \label{fig:WISCgaussesresult}
\end{figure}

The influence of artificially boosting redshift uncertainties for the Gaussian model is shown in Figure \ref{fig:WISCgaussesresult}. When increasing the scatter $\sigma$, not only does the posterior of $H_0$ widen, but also the peak of the distribution shifts. For the $2\sigma$ case the peak moves to the left, because the distribution of observed redshifts shifts to lower values as the uncertainties increase, as shown if Figure \ref{fig:mockcat}. However, for the $5\sigma$ case the distribution moves to the right, and overlaps with the empty catalog case. This is because for such large uncertainties (amounting to between 0.1 up to 0.25) the catalogue is uninformative and the posterior of $H_0$ is dominated by the out-of-catalogue part, and is pushed towards the empty catalogue case.

\begin{figure}
    \centering
    \includegraphics[width=0.5\textwidth]{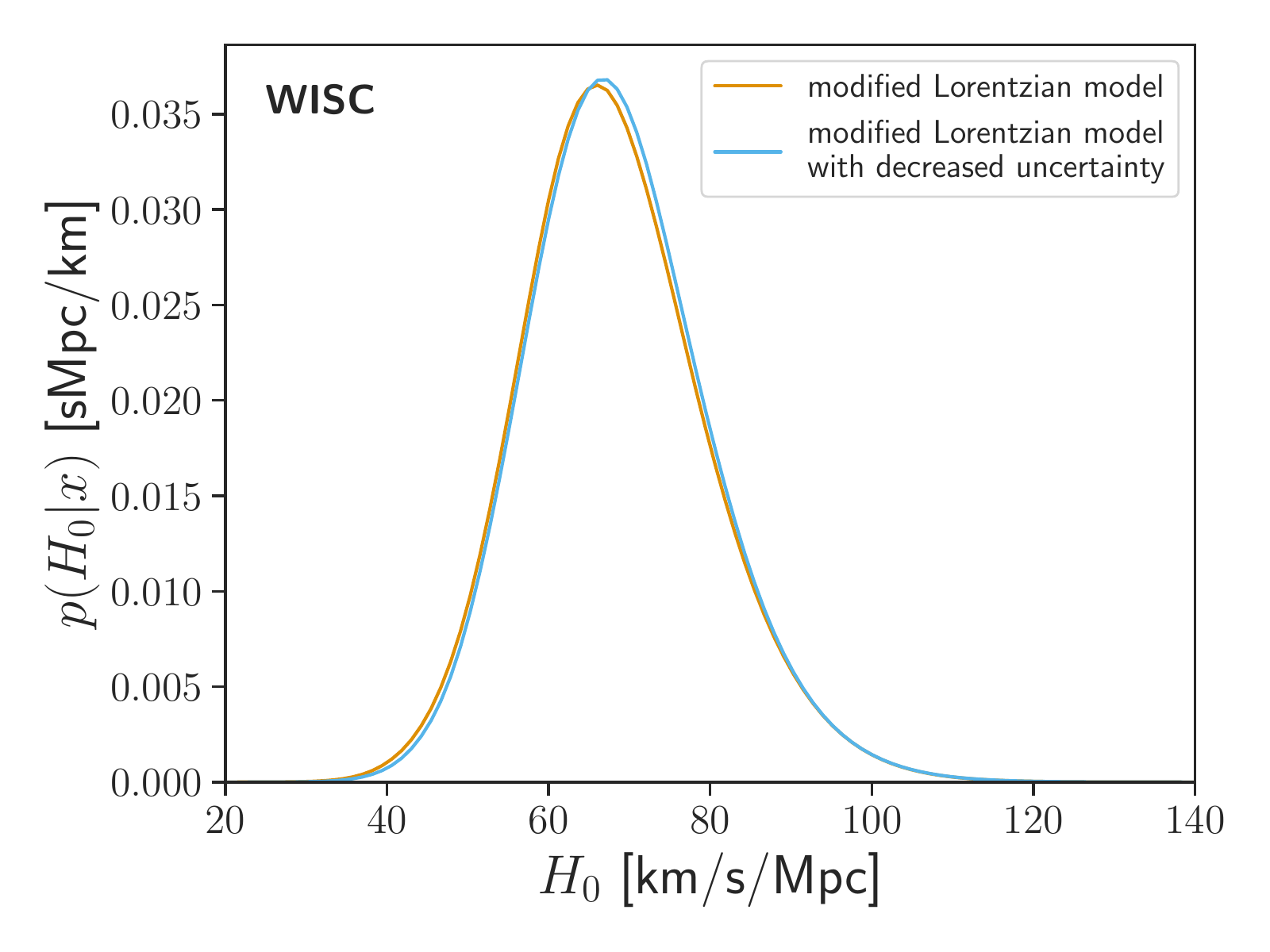}
    \caption{The influence of decreasing the redshift uncertainty in the modified Lorentzian model for the WISC catalogue. We plot the $H_0$ posterior density with the best-fit modified Lorentzian model (blue), and the same model but with uncertainties taken for galaxies at twice smaller redshift (orange; see details in the text). This change does not have a significant impact on the $H_0$ posterior.}
    \label{fig:WISCmlorentzesresult}
\end{figure}

The last case we studied was to artificially change the modified Lorentzian uncertainties. This
is not as straightforward as for the Gaussian, because the model given by Eq.~(\ref{eq:mlorentz}) puts a limit on the parameter $a>0.5$. Therefore,  the best-fit model for WISC is valid only when $z_\text{photo}<0.424$. Moreover, the parameters $s$ and $a$ are not independent and both are responsible for the shape of the distribution. Hence, to preserve the realistic shape of the error distribution, instead of varying $a$ and $s$ directly, we modify the model by using for a galaxy with  $z_\text{photo}$, the redshift error value that would be assigned to an object with $0.5*z_\text{photo}$, 
effectively choosing the error of twice as small redshift. The result is shown in Figure \ref{fig:WISCmlorentzesresult}. Decreasing the uncertainty this way has a minimal effect on the shape of the $H_0$ posterior. 

\begin{figure}
    \centering
    \includegraphics[width=0.5\textwidth]{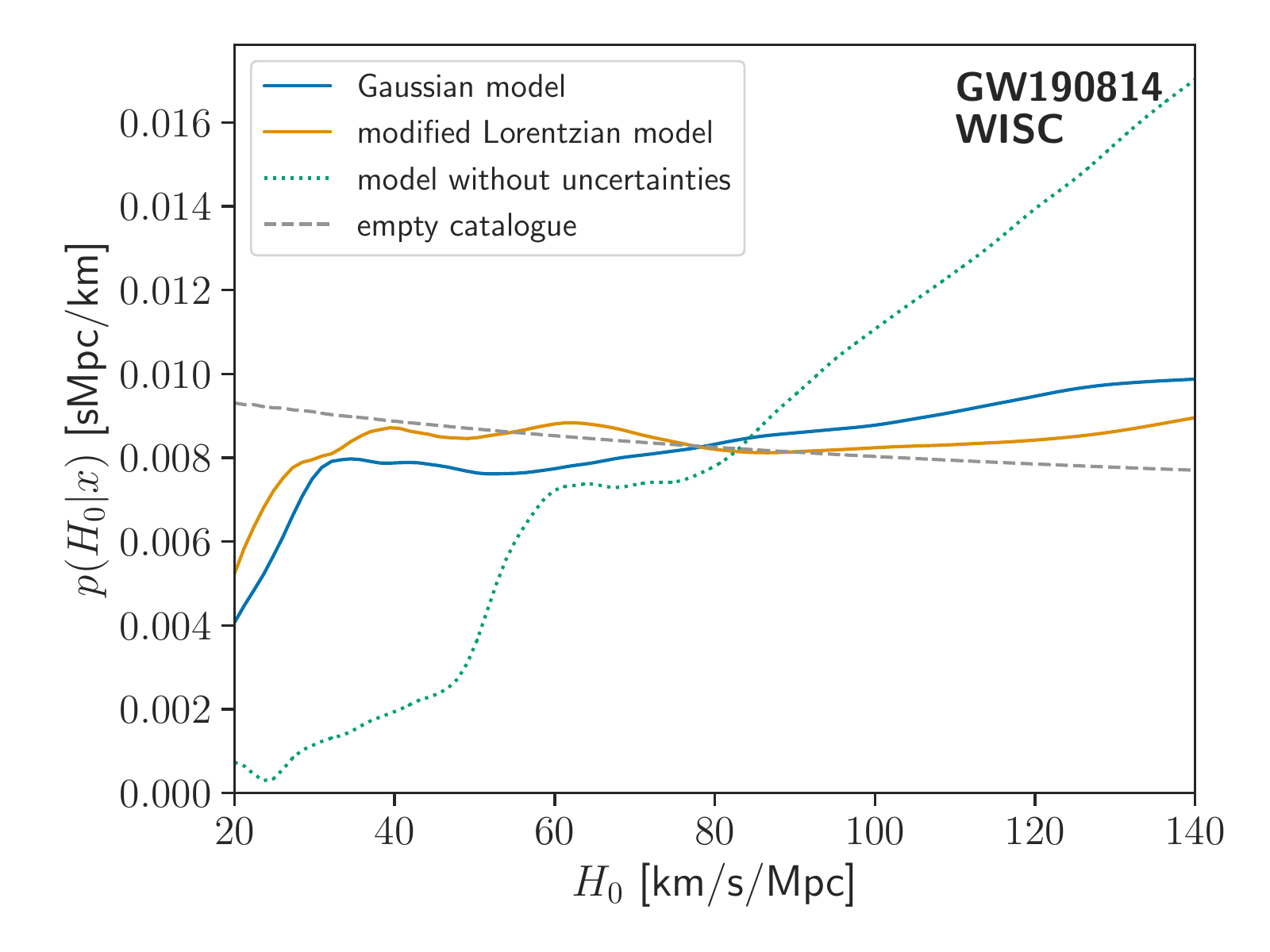}
    \caption{The influence of redshift uncertainties on the $H_0$ obtained from the single well-localized event GW190814 which has a significant in-galaxy-catalog contribution. In this case, we see a visible difference between the result with Gaussian and modified-Lorentzian uncertainties and a significant change when redshift uncertainties are turned off altogether. We would like to emphasise that the result obtained without redshift uncertainties is not to be trusted due to reasons explained in the text.}
    \label{fig:GW190814_impact}
\end{figure}

To conclude this section, we study the above effects focusing on a single GW detection, GW190814, and the WISC catalog. GW190814 is the best-localized dark standard siren to date, and it was seen in \citet{LVK_H0} that it is one of the few events in GWTC-3 with significant in-catalog information. The posterior distributions on $H_0$ with Gaussian, modified-Lorentzian, and without redshift uncertainties are shown in Figure~\ref{fig:GW190814_impact}. In this case, we see a significant difference with and without redshift uncertainties and also a more visible difference coming from different uncertainty profiles. Turning off the redshift uncertainties altogether leads to $H_0$ peaking at higher values. This is explained by an overdensity of galaxies along the line of sight of GW190814 seen at a redshift of around $0.22$ when redshift uncertainties are not considered.  We discuss this further in Appendix~\ref{sec:app-GW190814}. Since this overdensity vanishes when redshift uncertainties are included, it is not a real feature in the data. These results further highlight the importance of including proper redshift uncertainties in a dark standard siren measurement of $H_0$.

\section{Conclusions and Outlook} \label{sec:conclusion}

In this paper, we looked into the impact of redshift uncertainties of potential host galaxies on a dark GW standard-siren measurement of $H_0$. We first constructed Gaussian and modified Lorentzian uncertainty models for two catalogues: WISC and 2MPZ. Then we modified the \textit{gwcosmo} code to handle the modified Lorentzian uncertainty model. We used the same set of 46 gravitational sources as in \citet{LVK_H0} to obtain the $H_0$ posterior with WISC and 2MPZ catalogues and with Gaussian, modified Lorentzian, and no uncertainty models of redshift uncertainties. 
Redshift uncertainties have an impact on the observed redshift distribution of galaxies in a catalogue and thus have an impact on the statistical inference of $H_0$. The impact is particularly large in the case of well-localized GW events with significant in-catalogue contributions such as GW190814. The impact however is not statistically significant with the current ensemble of GW detections. It is worth highlighting that the impact on $H_0$ of different redshift uncertainty profiles (as seen in Figure~\ref{fig:WISCresult}, for example) is slightly smaller, but of the same order of magnitude as some of the principal sources of systematic errors in \cite{LVK_H0}, namely the uncertainty in compact binary population model parameters (Figure~11 of \citealt{LVK_H0}). It is to be noted that while this work was in preparation, two separate methods, \cite{Mastrogiovanni:2023emh,Mastrogiovanni:2023zbw} and \cite{Gray:2023wgj}, were developed for marginalizing over the unknown parameters of the population model, potentially making uncertainties coming from the galaxy catalogue sector the leading source of possible systematic errors in the dark standard siren $H_0$ measurement.

The relative importance of redshift uncertainties is expected to increase in the course of the future observing runs as more GW events come in and the statistical measurement errors are reduced. Moreover, uncertainties coming from the galaxy catalogue sector are expected to become more important with future deeper galaxy surveys. We recommend a follow-up mock data challenge in order to investigate various sources of systematic uncertainties important for the fourth observing run of the LVK detector network. Along with other objectives, we expect this mock data challenge to also tell us the exact conditions when redshift uncertainty models will become significantly important.

\section*{Acknowledgements}

We would like to thank Tessa Baker, Freija Beirnaert, Martin Hendry, Christos Karathanasis, Simone Mastrogiovanni, Surhud More, Suvodip Mukherjee, Federico Stachurski, and Nicola Tamanini for fruitful discussions throughout the project, and additionally Stefano Rinaldi and Aditya Vijaykumar for useful comments on the manuscript. We would also like to thank the anonymous referee for suggesting additional analysis which strengthened our conclusions.

The research of CT, GD, and AG is supported by the Ghent University Special Research Funds (BOF) project BOF/STA/202009/040 and the Fonds Wetenschappelijk Onderzoek (FWO) iBOF project BOF20/IBF/124. The research of RG is supported by the European Research Council, starting grant SHADE 949572.
MB is supported by the Polish National Science Center through grants no. 2020/38/E/ST9/00395, 2018/30/E/ST9/00698, 2018/31/G/ST9/03388 and 2020/39/B/ST9/03494, and by the Polish Ministry of Science and Higher Education through grant DIR/WK/2018/12.

This material is based upon work supported by NSF's LIGO Laboratory which is a major facility fully funded by the National Science Foundation. The computing runs for the cosmology inference using \texttt{gwcosmo} have been performed on the LIGO Data Grid computer clusters. The authors are grateful for computational resources provided by the LIGO Laboratory and supported by National Science Foundation Grants PHY-0757058 and PHY-0823459. This work makes use of gwcosmo which is available at \url{https://git.ligo.org/lscsoft/gwcosmo}.

This manuscript was reviewed by the LIGO Scientific Collaboration (document number: LIGO-P2300044) and by the Virgo Collaboration (document number: VIR-0147A-23).

\section*{Data Availability}

The 2MPZ and the WISC catalogues are available at \href{http://ssa.roe.ac.uk/TWOMPZ.html}{http://ssa.roe.ac.uk/TWOMPZ.html}, and at \href{http://ssa.roe.ac.uk/WISExSCOS.html}{http://ssa.roe.ac.uk/WISExSCOS.html}, respectively. All of the GW events we have used in this analysis are available at \href{https://www.gw-openscience.org/}{https://www.gw-openscience.org/}.

\appendix
\section{Redshift uncertainty model calibration}
\label{app:model}

As we do not have individual photo-z PDFs nor photo-$z$ uncertainties for each photometric galaxy, we use a calibration spectroscopic sample to build a photo-$z$ error model for the entire photometric sample. To go beyond the 0\textsuperscript{th} order approximation where this model would be the same for all the galaxies (i.e. independent of any observable quantity), we assume that the shape of the photo-$z$ PDF will depend on the photo-$z$ itself (it could instead, or also, depend for instance on galaxy magnitude or color). We could then fit a linear (or any other) relation between the photo-$z$ and its mean error based on the individual $z_\text{phot}-z_\text{spec}$ residuals. However, we want to model also the scatter (standard deviation in the Gaussian case, or more generally the wings of the modified Lorentzian distribution), which requires some binning. For good statistics we choose relatively broad bins in photo-$z$ and the parameters of the model are derived for these bins, but still as simple functions of photo-$z$. This allows us to build generalised models as a function of redshift.

\begin{figure*}
    \centering
    \includegraphics[width=0.4\textwidth]{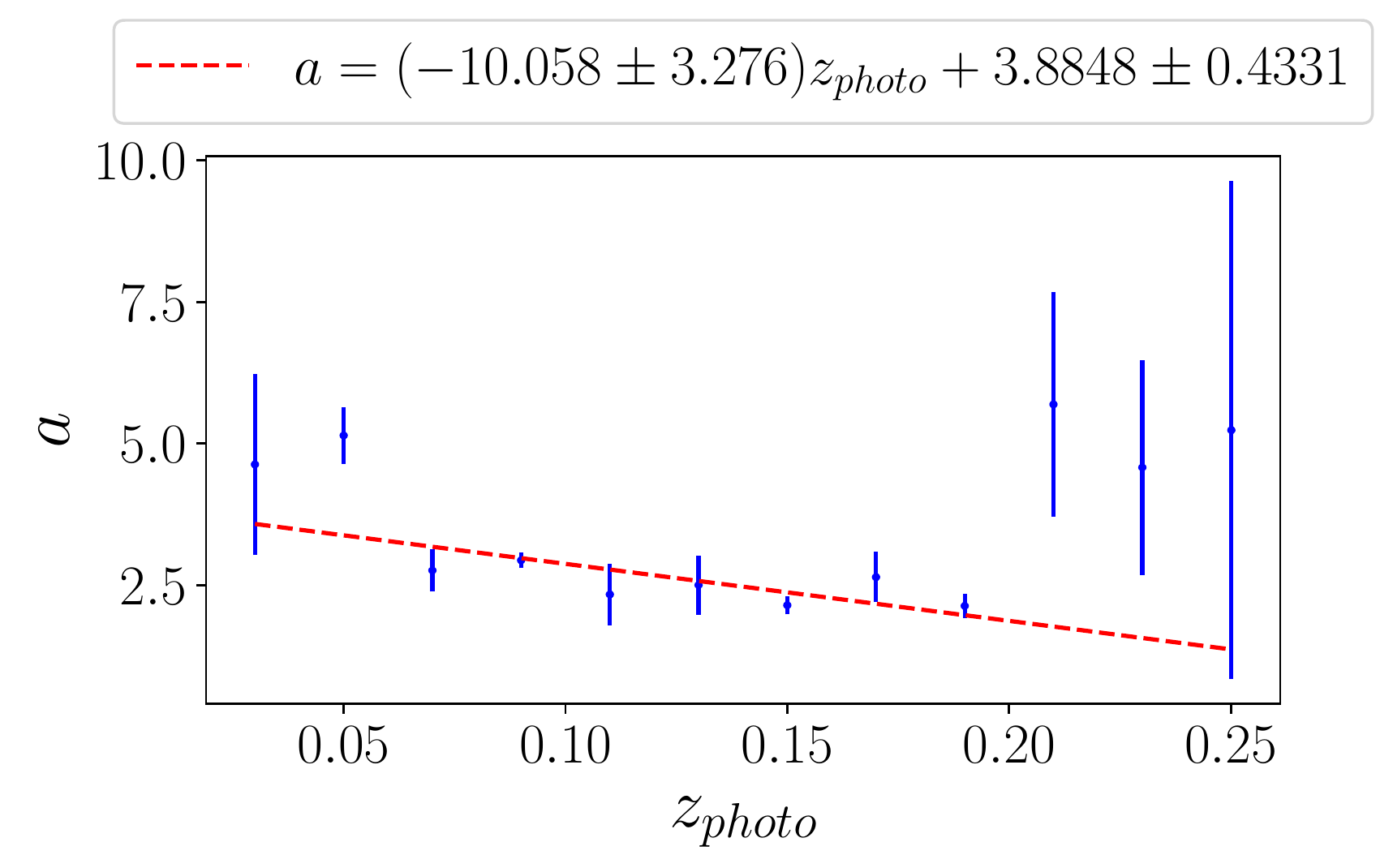}
    \includegraphics[width=0.4\textwidth]{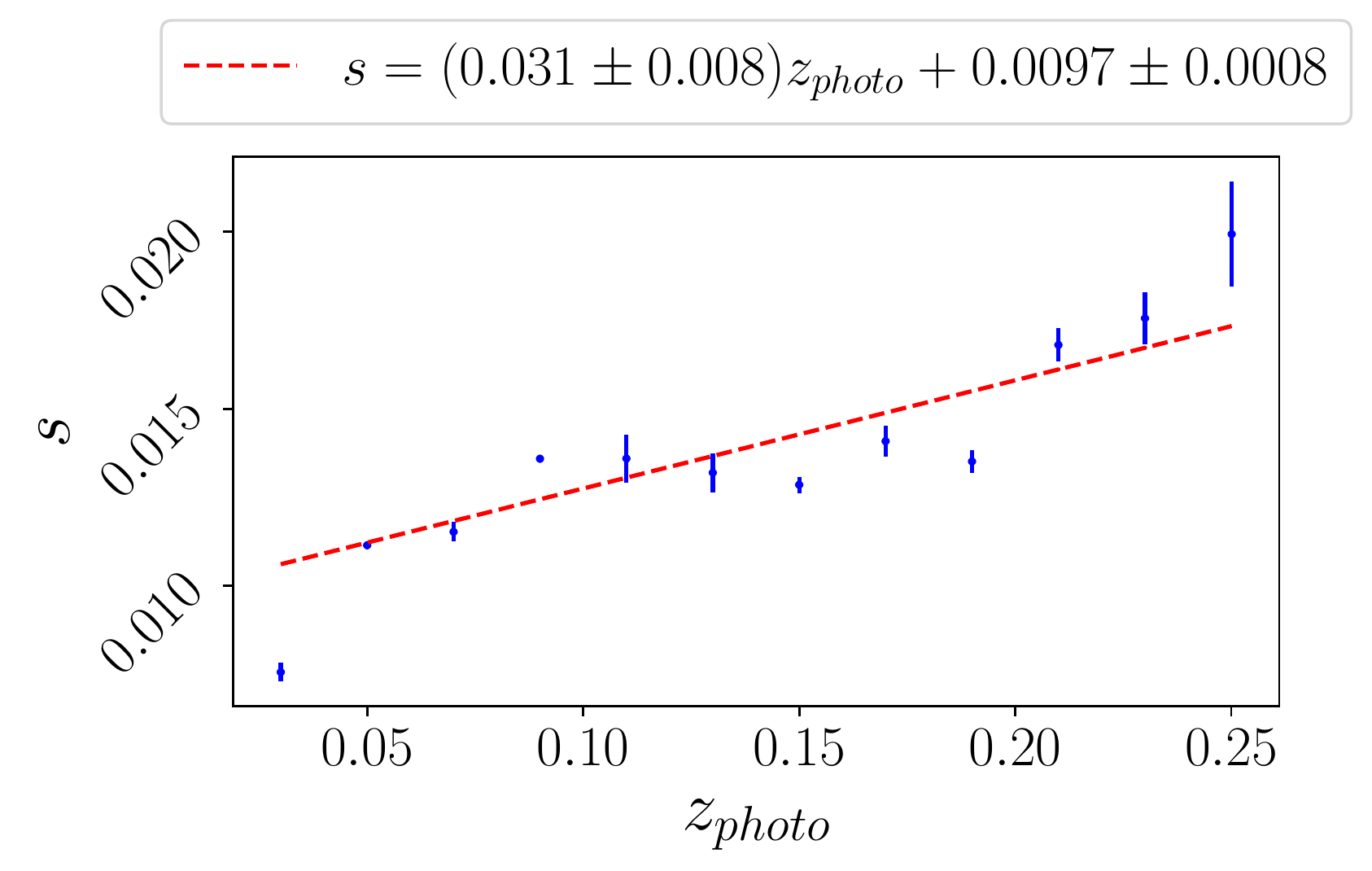}
    \includegraphics[width=0.4\textwidth]{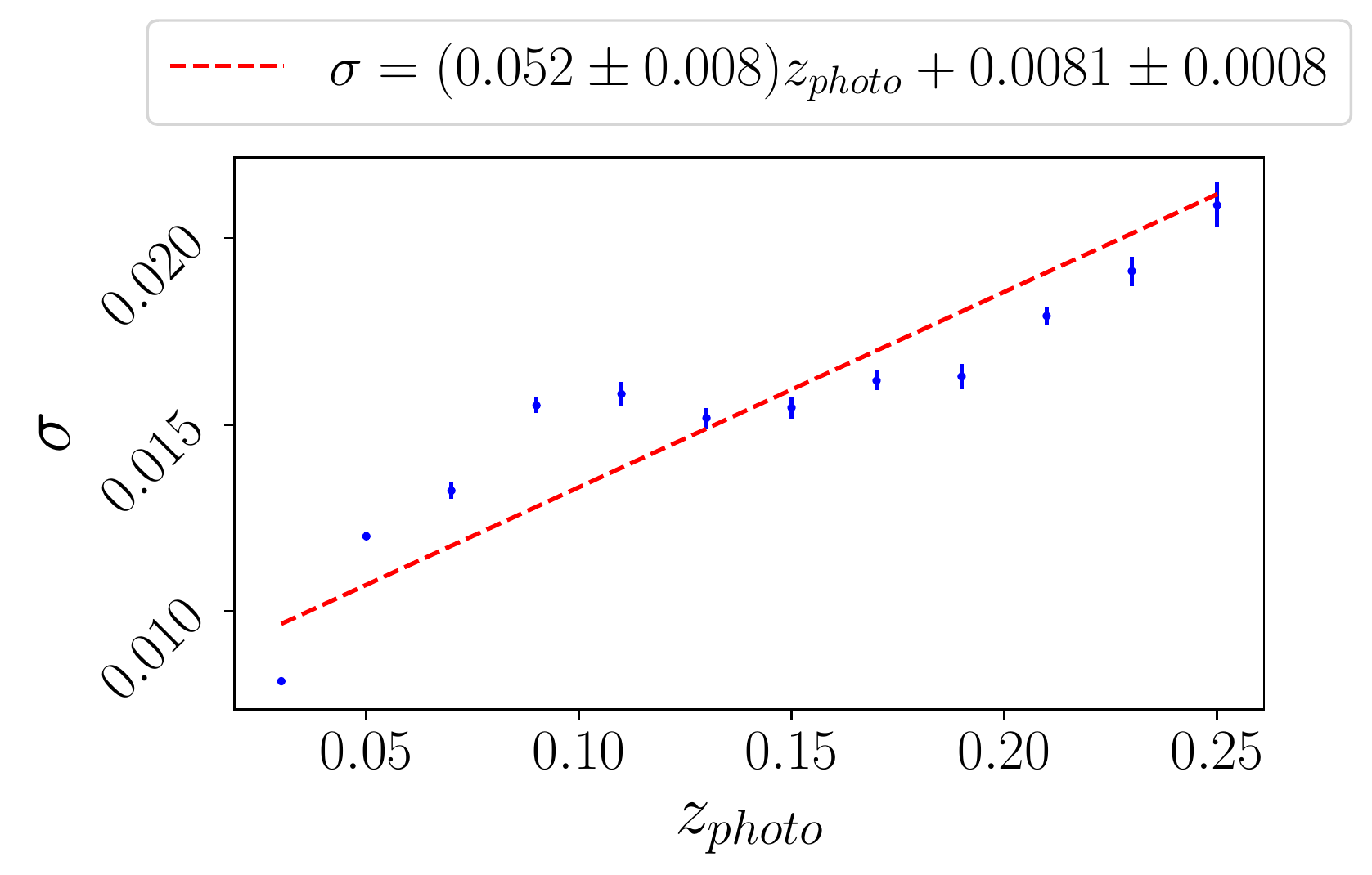}
    \includegraphics[width=0.4\textwidth]{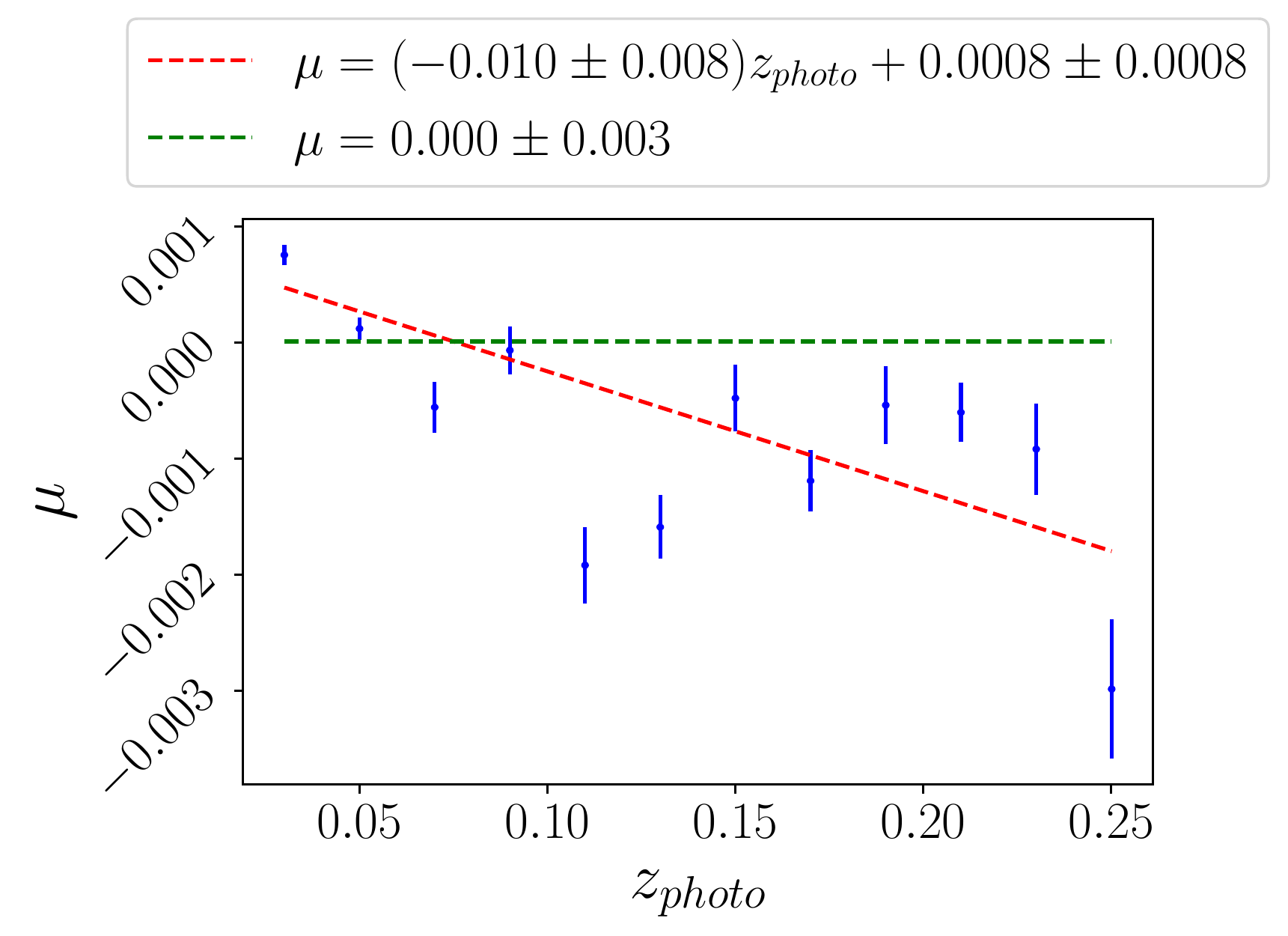}
    \caption{Models of parameters $a$, $s$ for modified Lorentzian model (top) and $\sigma$, $\mu$ for Gaussian model (bottom) obtained for 2MPZ from the calibration sample. Points represent parameters obtained from each bin and the middle of each bin. We use $\mu=0$ for our further calculations.}
    \label{fig:2MPZ_params}
\end{figure*}

\begin{figure*}
    \centering
    \includegraphics[width=0.4\textwidth]{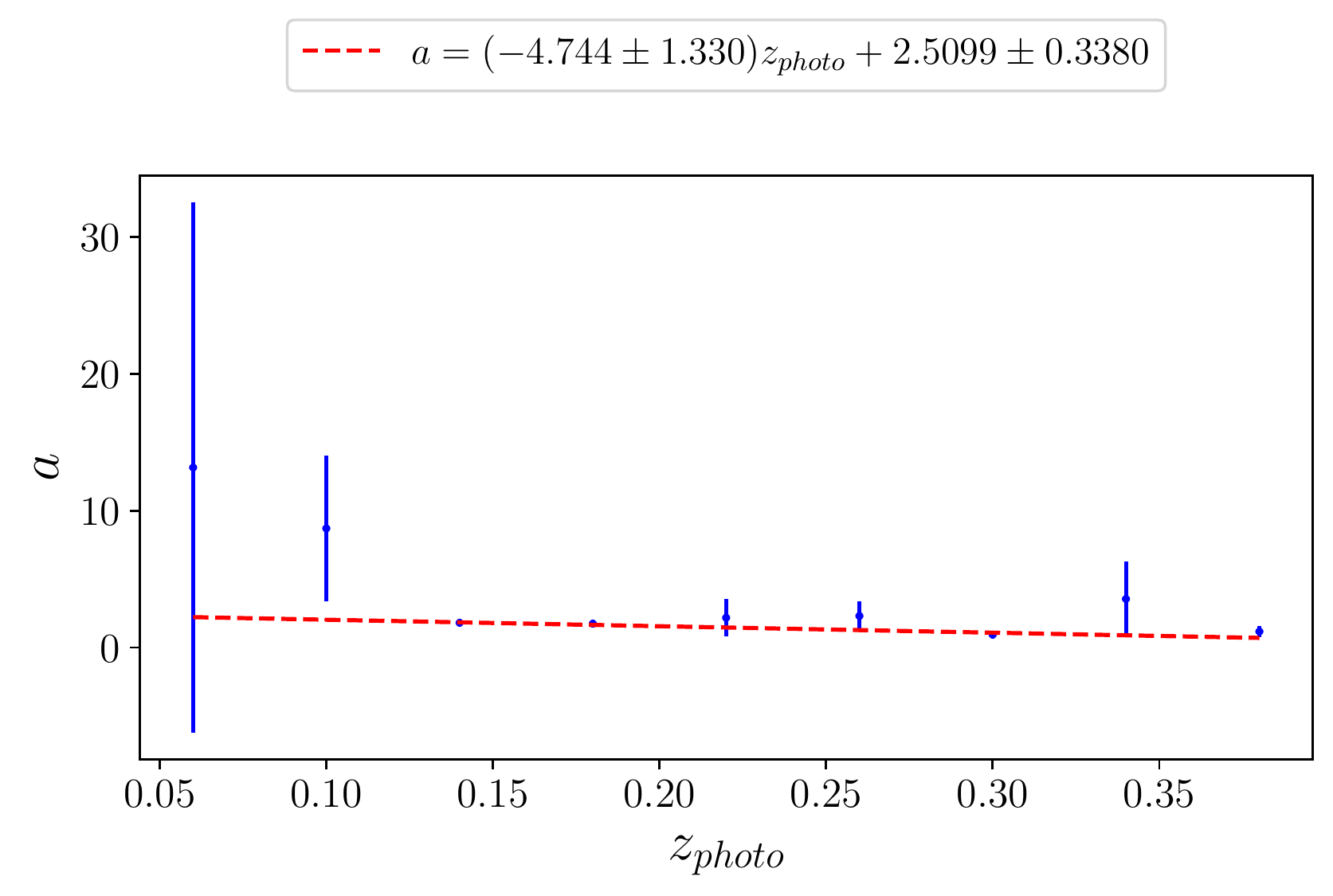}
    \includegraphics[width=0.4\textwidth]{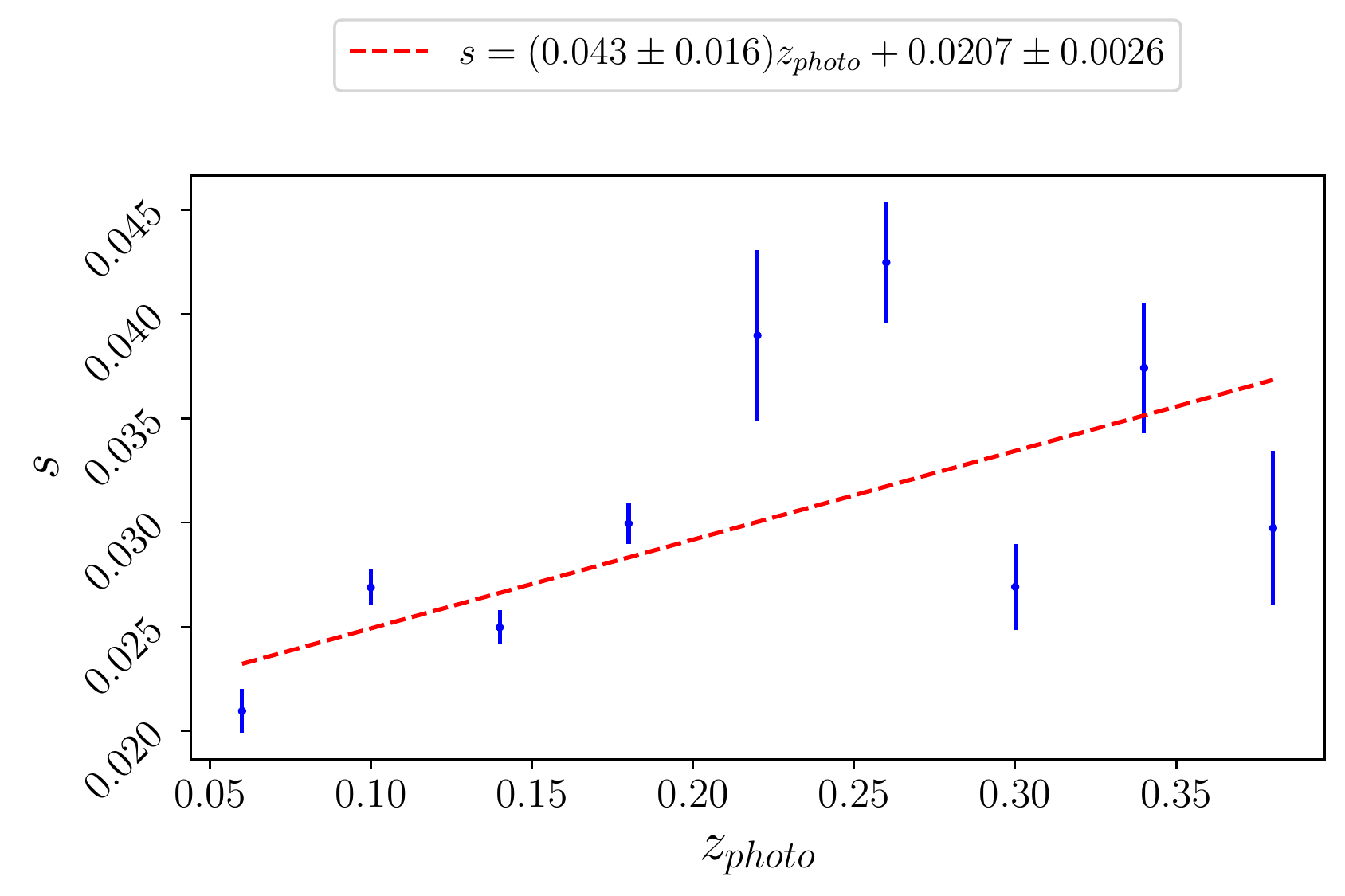}
    \includegraphics[width=0.4\textwidth]{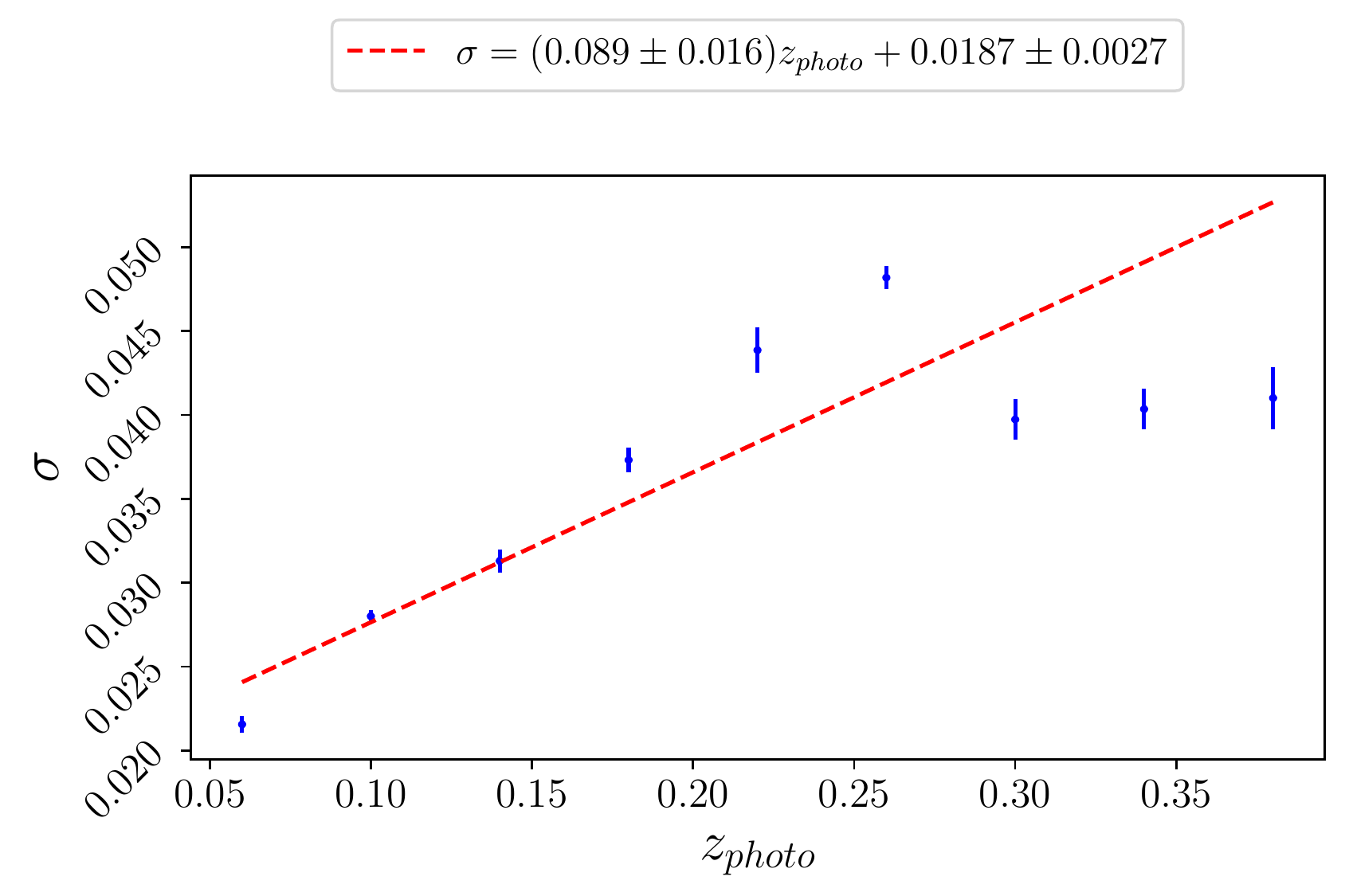}
    \includegraphics[width=0.4\textwidth]{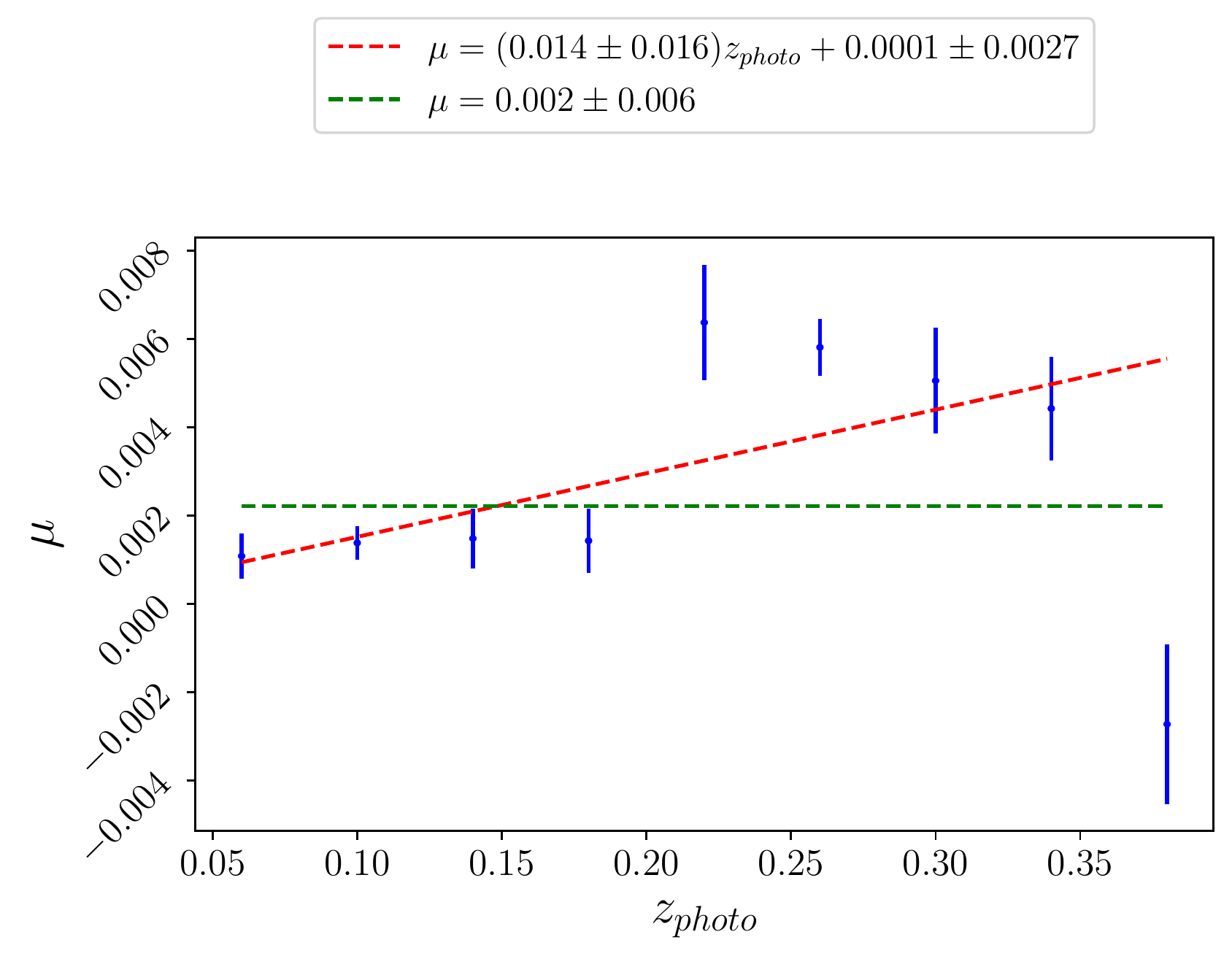}
    \caption{Models of parameters $a$, $s$ for modified Lorentzian model (top) and $\sigma$, $\mu$ for Gaussian model (bottom) obtained for WISC from the calibration sample. Points represent parameters obtained from each bin and the middle of each bin. We use $\mu=0$ for our further calculations.}
    \label{fig:WISC_params}
\end{figure*}

We consider Gaussian and modified-Lorentzian uncertainty models with parameters described in the main text in Section~\ref{sec:redshift}.
We assume linear dependence of each parameter and photometric redshift then we fit linear function for 2MPZ and WISC. Results are shown in Figure \ref{fig:2MPZ_params} and \ref{fig:WISC_params} respectively. 

\section{Galaxy overdensities along the line-of-sight of GW190814}
\label{sec:app-GW190814}

In order to investigate the behaviour seen in Figure~\ref{fig:GW190814_impact}, we look at the redshift distribution of galaxies in the WISC catalogue along the line-of-sight of GW190814 with different uncertainty models. We find an overdensity around $z=0.24$ when redshift uncertainties are turned off, which is extraneous and goes away as soon as these uncertainties are taken into account. This explains why $H_0$ tend to favour higher values in the absence of redshift uncertainties in Figure~\ref{fig:GW190814_impact}.

\begin{figure}
    \centering
    \includegraphics[width=0.5\textwidth]{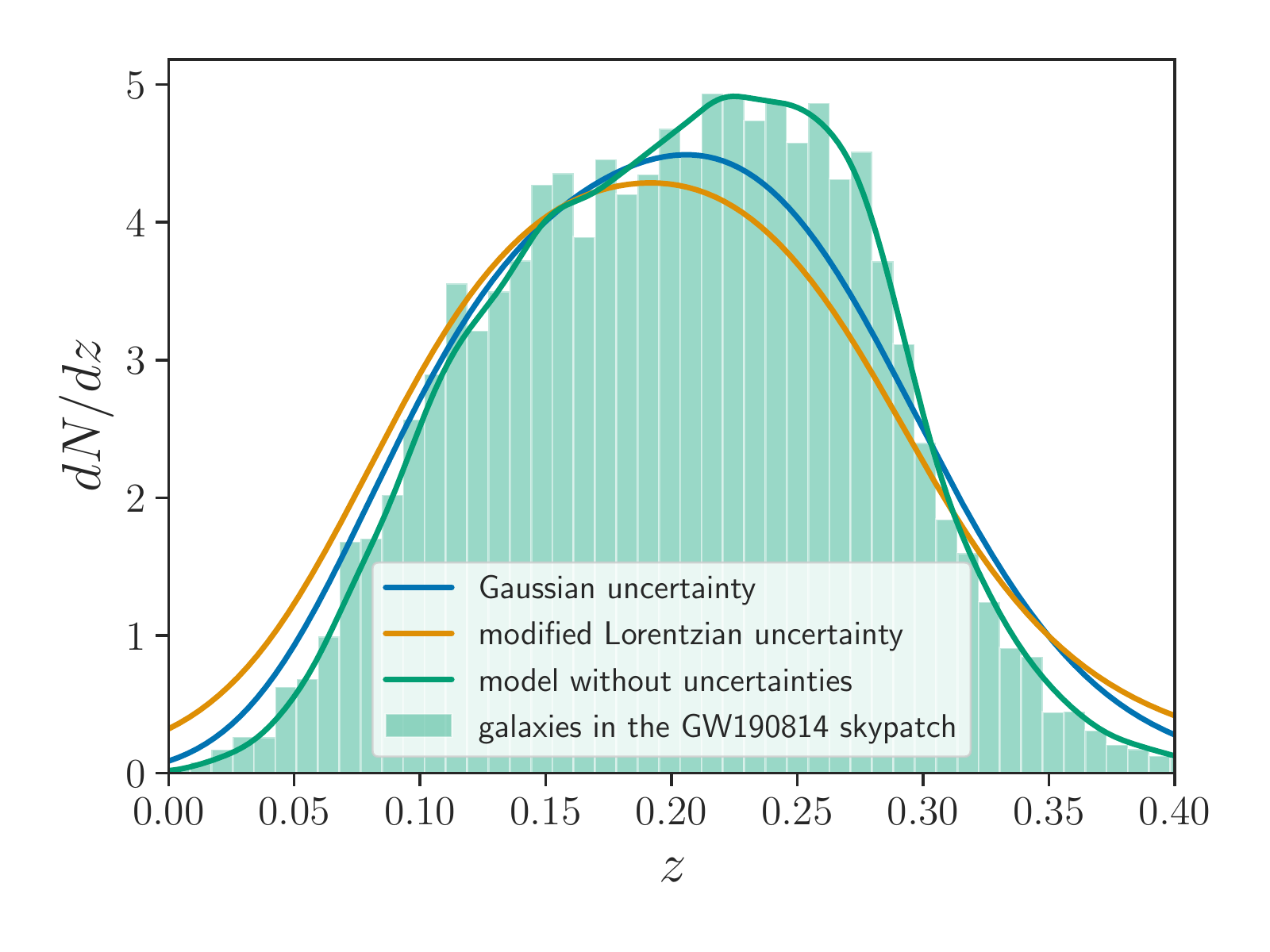}
    \caption{The redshift distribution from galaxies from the WISC catalogue along the line-of-sight of GW190814. The 90\%~sky area is considered. As in the rest of the manuscript, we consider the Gaussian and Lorentzian redshift uncertainty models, and also no redshift uncertainties. In the absence of redshift uncertainties, we see an overdensity of galaxies around $z=0.24$, which goes away when uncertainties are taken into account.}
    \label{fig:GW190814_los}
\end{figure}



\bibliographystyle{mnras}
\bibliography{z_uncert_main}


\bsp
\label{lastpage}
\end{document}